\documentclass{openjournal}

\usepackage{lipsum}

\usepackage{xcolor}
\usepackage{textgreek}
\usepackage[utf8]{inputenc}
\usepackage[english]{babel}
\usepackage{float}

\definecolor{linkcolor}{rgb}{0.0,0.3,0.5}
\usepackage{hyperref}
\hypersetup{
    unicode, 
    colorlinks=true,
    linkcolor=blue,
    citecolor=blue,
    filecolor=blue,
    urlcolor=blue,
}

\def\cdl#1{\textcolor{black}{#1}} 
\def\mmt#1{\textcolor{black}{#1}} 

\usepackage{tensind}
\tensordelimiter{?}
\DeclareGraphicsExtensions{.bmp,.png,.jpg,.pdf}
\usepackage{verbatim}
\usepackage[normalem]{ulem}
\usepackage{orcidlink}
\usepackage{soul}
\usepackage{amsmath}

\graphicspath{ {./figs/} }

\begin{document}
\title{Current and future constraints on the expansion history of the GREA model}

\author{Irene Graziotti\orcidlink{0009-0003-3026-826X}}
\email{irene.graziotti@inaf.it}
\affiliation{INAF-Osservatorio Astronomico di Capodimonte, Via Moiariello 16, Napoli, 80131, Italy}

\author{Chiara De Leo\orcidlink{0009-0009-1175-213X}}
\email{chiara.deleo@uniroma1.it}
\affiliation{Physics Department, Sapienza University of Rome, P.le A. Moro 5, 00185 Roma, Italy}
\affiliation{INAF - Osservatorio Astronomico di Roma, via Frascati 33, 00078 Monte Porzio Catone, Italy}
\affiliation{Istituto Nazionale di Fisica Nucleare (INFN), Sezione di Roma, P.le A. Moro 5, I-00185, Roma, Italy}
\affiliation{Physics Department, Tor Vergata University of Rome, Via della Ricerca Scientifica 1, Roma, 00133, Italy}

\author{Matteo Martinelli\orcidlink{0000-0002-6943-7732}}
\email{matteo.martinelli@inaf.it}
\affiliation{INAF - Osservatorio Astronomico di Roma, via Frascati 33, 00078 Monte Porzio Catone, Italy}
\affiliation{Istituto Nazionale di Fisica Nucleare (INFN), Sezione di Roma, P.le A. Moro 5, I-00185, Roma, Italy}

\begin{abstract}
In this work, we investigate the General Relativistic Entropic Acceleration (GREA) framework, in which late-time acceleration emerges from entropy production associated with the cosmological horizon, and compare its performance with the standard $\Lambda$CDM description of the Universe.

We first confront GREA with current background observations, including baryon acoustic oscillations, type Ia supernovae, compressed CMB information, and cosmic chronometers, with particular emphasis on the geometric horizon parameter $\sqrt{-k}\eta_0$.
We then introduce a phenomenological extension of the theory by allowing for an additional dark energy component, $\Omega_{de}$, enabling the recovery of a $\Lambda$CDM-like expansion history as a limiting case.

We perform a Bayesian parameter inference and model comparison analysis using both current data and mock datasets representative of future surveys, including SKAO, LSST, and ET.
While current data statistically prefer $\Lambda$CDM when compressed CMB information is included, GREA remains competitive for low-redshift combinations.
Forecasts indicate that gravitational wave standard sirens are expected to enhance the ability to discriminate between entropic-driven and dark-energy-driven expansion scenarios, and to identify the underlying cosmological model favored by the data.

\end{abstract}


\maketitle

\section{Introduction}
\label{sec:intro}
The discovery of the accelerated expansion of the Universe represents one of the most profound challenges in modern cosmology.
Observations of type Ia supernovae (SNIa) \citep{Riess_1998}, together with measurements of the cosmic microwave background (CMB) \citep{Plank_2018} and large-scale structure (LSS) \citep{sdsscollaboration2025nineteenthdatareleasesloan}, consistently indicate that the current expansion of the Universe is dominated by a dark energy component with negative pressure.  

Within the standard cosmological model, also known as the $\Lambda$CDM model, this accelerated expansion is attributed to the cosmological constant $\Lambda$.
Over the past decades, an increasingly wide range of high-precision cosmological observations has provided strong and independent support for this framework.
Current measurements, analyzed within the context of $\Lambda$CDM, indicate an Universe with an energy budget currently composed of approximately $5\%$ baryonic matter, $25\%$ cold dark matter, $70\%$ dark energy (identified in $\Lambda$), with minor contributions from massive neutrinos and radiation \citep{Adame_2025}.

Despite its remarkable agreement with a wide range of cosmological observations, the $\Lambda$CDM model faces several theoretical and observational issues.
From a theoretical perspective, these include the cosmological constant problem, related to the discrepancy between the observed value of $\Lambda$ and vacuum energy estimates from quantum field theory \citep{YaBZel'dovich_1968}, as well as the coincidence \citep{Zlatev:1998tr,Velten_2014} and fine-tuning \citep{devuyst2024naturalintroductionfinetuning} problems.
On the observational side, several tensions have emerged between early- and late-time measurements, most notably the Hubble tension \citep{Abdalla_2022, Riess_2022}, which highlights a significant discrepancy between the value of the Hubble constant, $H_0$, inferred from CMB and that measured from the local Universe.
Additional problems, including the nature of dark matter \citep{Weinberg_2015} and anomalies such as the Lithium-7 problem \citep{Mathews_2017}, further motivate the exploration of alternative cosmological scenarios beyond $\Lambda$CDM.
These challenges suggest that the late-time accelerated expansion of the Universe may not be fully captured by a simple cosmological constant, but may instead arise from a fundamental dark energy component or from new physics beyond the standard cosmological model.

These considerations have led to an extensive investigation of alternative cosmological models.
One class of such scenarios involves dynamical dark energy models, where the accelerated expansion is driven by an evolving scalar field.
In these frameworks, the dark energy equation of state varies with cosmic time, modifying the expansion history of the Universe.
A useful distinction can be made between early dark energy and late dark energy models.
Early dark energy scenarios \citep{Di_Valentino_2021} introduce a dark energy component in the early Universe, affecting the physics of recombination and leaving imprints on the cosmic microwave background.
Late dark energy models \citep{Di_Valentino_2021}, on the other hand, assume that dark energy becomes dynamically relevant only at relatively recent epochs, modifying the expansion history at low redshifts.
An alternative and complementary approach is provided by modified gravity theories \citep{Bahamonde_2018}, which explain the accelerated expansion through the intrinsic dynamics of spacetime geometry without invoking a cosmological constant.

Within this broad landscape of alternatives, the Gravitational Entropic Acceleration (GREA) framework provides a novel perspective on cosmic acceleration \citep{ Garc_a_Bellido_2021, Arjona_2022, garciabellido2024darkenergypredictionsgrea,  calderon2025constraininggreaalternativetheory, garciabellido2025greadarkenergyholographic}.
In GREA, the contribution to the Friedmann equations normally associated with dark energy emerges from a thermodynamic and entropic interpretation of gravity, rather than from the introduction of a fundamental dark energy component.
As a result, cosmic acceleration arises as an emergent phenomenon linked to horizon thermodynamics rather than a fundamental fluid.
Compared to $\Lambda$CDM, GREA predicts a modified expansion history while retaining a minimal set of additional assumptions, making it a theoretically appealing and observationally testable alternative.
However, the viability of such a framework must be carefully assessed through a detailed comparison with observational data.

In this work, we test the GREA theory with current observational data, including BAO, SNIa, CMB and cosmic chronometers (CC), to validate our implementation against previous studies and to assess the performance of the model. 

Furthermore, we extend the GREA framework by introducing a modified version of the model. 
We introduce an additional component to extend its phenomenology, which increases the number of degrees of freedom, potentially improving the agreement with observational data and offering further insight into the nature of the acceleration of late-time Universe.
Crucially, this modification is constructed so that the standard $\Lambda$CDM model is recovered as a limiting case.

The main objective of this article is to test GREA and its modified extension against current cosmological observations and to assess their performance relative to the standard $\Lambda$CDM model.
We analyze a combination of observational probes, which together provide complementary constraints on the expansion history of the Universe.
In addition to the analysis of current data, a key goal of this work is to investigate the constraining power of future observations and to assess whether next-generation surveys will be able to discriminate between $\Lambda$CDM, GREA and its modified extension.
For this purpose, we perform a forecast analysis assuming different fiducial cosmologies.
Thanks to this analysis, we test the internal consistency of each model and evaluate their ability to recover the underlying cosmology.

The paper is organized as follows: \autoref{sec:GREA_theory} introduces the GREA cosmological framework, derives the main equations relevant for our study, and explains the modified GREA theory; \autoref{sec:dataset} describes the observational datasets and mock catalogs used in the analysis; \autoref{sec:analysis} presents the statistical framework, likelihood construction, and parameter inference.
The resulting constraints on cosmological parameters and model comparisons are discussed in \autoref{sec:results}.
We conclude in \autoref{sec:conclusion} with a summary of the main results and a discussion of possible future developments.

\section{General Relativistic Entropic Acceleration (GREA) theory}
\label{sec:GREA_theory}

We review the main equations of the General Relativistic Entropic Acceleration (GREA), following the original work of \citet{Garc_a_Bellido_2021}.

The GREA theory offers an alternative explanation to the standard cosmological model by accounting for the late-time accelerated expansion of the Universe without invoking a cosmological constant or other forms of dark energy.

In this framework, the cosmic acceleration is interpreted as an entropic force originating from the boundary entropy of the cosmological horizon. 
This force is negligible in the early Universe but becomes relevant at late times as the horizon grows.
At the present cosmological epoch, the entropy growth of this horizon leads to an effective repulsive contribution that counteracts the gravitational attraction of matter, determining the observed late-time accelerated expansion.

Within this approach, the Einstein field equations, extended to out-of-equilibrium phenomena, are modified with an additional term, the entropic force tensor $f_{\mu\nu}$, as \citep{Arjona_2022}:
\begin{equation}
\label{eq:einstein_mod}
R_{\mu\nu}-\frac{1}{2}Rg_{\mu\nu} = \frac{8 \pi G}{c^4}(T_{\mu\nu}-f_{\mu\nu}).
\end{equation}
The presence of $f_{\mu\nu}$ comes from the first law of thermodynamics and introduces an effective negative pressure associated with entropy production, $p_S=-TdS/dV<0$, according to the second law of thermodynamics.
Consequently, the energy conservation equation is modified to:
\begin{equation}
\label{eq:energycons}
\dot{\rho}+3H(\rho+p)=\frac{T\dot{S}}{a^3},
\end{equation}
where $S$ is the entropy per comoving volume, $T$ denotes the temperature, $\rho$ is the energy density and $H=\dot{a}/a$ is the Hubble parameter, and the dot denotes differentiation with respect to cosmic time.
This approach is grounded in an extension of the Einstein-Hilbert action with a surface term, the Gibbons-Hawking-York term (GHY) \citep{PhysRevD.15.2752}.
This allows for a temperature, $T_H$, and entropy, $S_H$, to be assigned to this horizon, that we can write as \citep{garciabellido2024darkenergypredictionsgrea, Espinosa_Portal_s_2021}:
\begin{equation}
\label{eq:TandS}
k_B T_H = \frac{\hbar}{2\pi} \frac{a ~\text{sinh}(2\eta\sqrt{-k})}{d_H^2\sqrt{-k}}, ~~~~~ S_H = \frac{k_B\pi}{\hbar} \frac{d_H^2}{G},
\end{equation}
where $d_H= a \eta$ is the causal cosmological horizon, $\eta$ is the conformal time and $k$ is the spatial curvature parameter.
The presence of $\hbar$ in \autoref{eq:TandS} should therefore be regarded as a distinctive feature of the nature of the horizon thermodynamics, where a quantum description of gravity is applied on a classical spacetime background.
So, in this case, the entropic contribution should be interpreted as an effective description of gravitational dynamics at large scales \citep{garciabellido2025greadarkenergyholographic}.

Our cosmological analysis focuses on a Universe composed of matter, radiation, and an entropic energy density term, $\rho_H=T_HS_H/a^3$, associated with the expansion of the horizon.
The total energy density is written as:
\begin{equation}
\label{eq:dentitytot}
    \rho=\rho_r+\rho_m+\rho_H.
\end{equation}
Although the total energy density is expressed as the sum of individual contributions, the different components are not independent.
The entropic energy density, $\rho_H$, arising from the GHY surface term \citep{Garc_a_Bellido_2021}, depends on the expansion history of the Universe through the evolution of the cosmological horizon, which is itself determined by the matter and radiation content.

A specific realization of this model is an open Universe in a phase of eternal inflation.

To determine the evolution of the scale factor, we solve the first Friedmann equation, expressed in terms of rescaled conformal time, $\tau=H_0 \eta = \int{dt/a(t)}$, following the work of \citet{calderon2025constraininggreaalternativetheory}:
\begin{equation}
\label{eq:grea}
    \tau^{\prime} = \frac{d\tau}{da} = \left[ a^2 \sqrt{\Omega_m a^{-3}\left(1 
+ \frac{a_{eq}}{a}\right) + \frac{4\pi}{3a^2}\frac{\text{sinh}(2\tau)}{(-k)^{3/2}V_c}}\right]^{-1},
\end{equation}
where $a_{eq}=\Omega_r/\Omega_m$ is the scale factor at the equality between radiation and matter, with $\Omega_r=\Omega_{\gamma}+\Omega_{\nu}$, (photons and massless neutrinos) and $\Omega_m=\Omega_b+\Omega_c$ (baryons and cold dark matter).
The comoving volume, $V_c$, is given by:
\begin{equation}
\label{eq:Vc}
V_c=\frac{\pi~\big[\text{sinh}(2\sqrt{-k}\eta_0)-2\sqrt{-k}\eta_0\big]}{(-k)^{3/2}}.
\end{equation}
The phenomenological impact of GREA theory is primarily determined by a single parameter, $\alpha$, which serves to characterize its late-time phenomenology and is defined as:
\begin{equation}
\label{eq:alpha}
    \alpha H(a_0) d_H(a_0) = \sqrt{-k}\eta_0,
\end{equation}
where the quantity $H(a) d_H(a)$ is the dimensionless horizon distance and $d_H(a)=a\eta(a)$ is the causal cosmological horizon.

It is important to emphasize that, within this formulation, the rescaled conformal time evaluated today, $\tau_0$, does not coincide with $H(a_0) d_H(a_0)$.
This reflects that the present-day expansion rate $H(z=0)$, obtained by numerically solving the GREA equation, generally does not coincide with the parameter $H_0$ used to define $\tau$, i.e. $H(z=0)\neq H_0$ \citep{garciabellido2024darkenergypredictionsgrea}.

In practice, $H_0$ is treated as an external parameter entering in the definition of the rescaled conformal time and in the normalization of the equation, while the physical expansion rate at the present epoch, $H(z=0)$, emerges as a derived quantity from the numerical solution of the modified Friedmann equation.
Although CMB observations constrain the early time expansion of the Universe, the present day expansion rate $H(z=0)$ is instead determined by the entropic contribution encoded in $\alpha$, or equivalently in $\sqrt{-k}\eta_0$. 
This leads to $H(z=0) \neq H_0$, even when the early time expansion history remains consistent with CMB constraints, allowing the present day expansion rate to be either larger or smaller than the value inferred under the $\Lambda$CDM assumption.
Further details on the numerical implementation of the difference between $H_0$ and $H(z=0)$ are described in \autoref{subsec:parameter_space}.
In the following part of the paper we will refer to $H(z=0)$ as $H(0)$. \mmt{It is important to stress once again that this quantity is the current expansion rate in the GREA model. For such a reason, this should be the quantity to be compared with independent measurements of the same physical quantities, e.g. with the distance ladder results, despite these quoting constraints on $H_0$.}

\begin{center}
    \begin{figure}[!t]
        \centering
    	\includegraphics[scale=0.3]{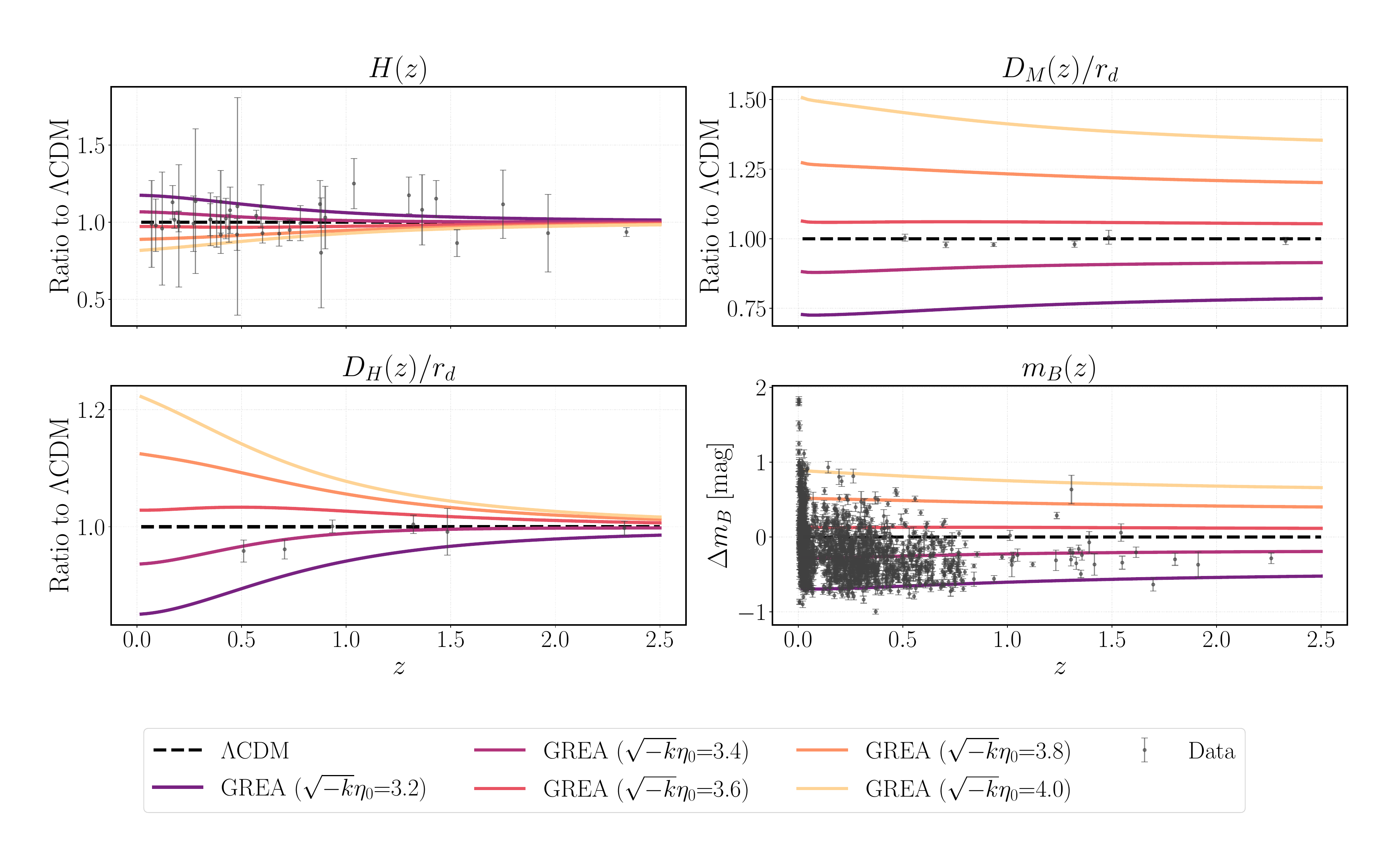}
    	\caption{\cdl{Ratios of $H(z)$ (top left), $D_M(z)/r_d$ (top right), and $D_H(z)/r_d$ (bottom left), together with the difference in apparent magnitude $\Delta m_B(z)$ (bottom right), evaluated in the GREA framework (solid lines) and normalized to $\Lambda$CDM (black dashed), for different values of $\sqrt{-k}\eta_0$. Observational data are shown in each panel: cosmic chronometers (top left) from \citet{Arjona_2022}, BAO (top right and bottom left) from DESI-DR2 \citep{Abdul_Karim_2025}, and Pantheon+ supernovae catalog (bottom right) \citep{Scolnic_2022}.}}
        \vspace{0.2 cm}
        \label{fig:H(z)GREALCDM}
    \end{figure}
\end{center}

In \autoref{fig:H(z)GREALCDM}, we present a comparison between $\Lambda$CDM and the GREA model for several cosmological observables, normalized to the standard scenario.
The modification primarily affects the late-time expansion, leading to deviations in $H(z)$, BAO distance measures, and supernova apparent magnitudes that increase with the parameter $\sqrt{-k}\eta_0$.
For GREA we solve \autoref{eq:grea} with initial conditions $a_i=10^{-11}$, $\tau_0=a_i/\sqrt{\Omega_r}$ and different values of $\sqrt{-k}\eta_0$. 

As depicted, the two models converge at early times at the level of the expansion rate $H(z)$, confirming that both theories accurately describe the dynamics of the matter dominated epoch.
In contrast, a substantial difference is observed for $z < 1$.
This difference is of particular interest, as it highlights how the distinct late-time dynamics of the two models leads to different predictions for the present-day expansion rate.

\subsection{GREA modified theory}
\label{subsec:GREAmod}
In addition to the standard GREA framework, we consider an empirical modification of the theory by introducing an additional dark energy contribution, analogous to a cosmological constant.

With this addition, \autoref{eq:grea} can be written as:
\begin{align}
 \label{eq:grea_mod}
        \tau^{\prime}  = \frac{d\tau}{da} = \left[ a^2 \sqrt{\Omega_m a^{-3}\left(1 
+ \frac{a_{eq}}{a}\right) + \frac{4\pi}{3a^2}\frac{\text{sinh}(2\tau)}{(-k)^{3/2}V_c} + \Omega_{de}}\right]^{-1}.
\end{align}

The additional term $\Omega_{de}$ describes a constant dark energy contribution within the modified GREA framework.
We emphasize that $\Omega_{de}$ is not derived from the entropic construction underlying GREA, but is introduced phenomenologically to test whether the data require this additional component in addition to the entropic effects.
This empirical extension allows the $\Lambda$CDM expansion history to be recovered as the limiting case, when the entropic contribution specific to GREA becomes negligible.

In particular, we can obtain the condition to reduce to the $\Lambda$CDM limit by substituting the expression of $V_c$ from \autoref{eq:Vc} in the entropic term:
\begin{equation}
\label{eq:lambdalimit}
\frac{\sinh(2\tau)}{(-k)^{3/2}V_c} \simeq \frac{\sinh{2\tau}}{\sinh(2\sqrt{-k}\eta_0)},
\end{equation}
which vanishes in the regime $\tau \ll \sqrt{-k}\eta_0$.
In this limit, the modified GREA dynamics reduces to that of $\Lambda$CDM.

It is important to stress that we do not impose spatial flatness, nor do we require the energy density parameters to satisfy the condition $\sum_i \Omega_i =1$.
The spatial curvature enters in the dynamics explicitly through the entropic term, with the curvature parameter $k$ and the comoving volume $V_c$, while the dark energy contribution $\Omega_{de}$ is treated as an independent parameter.
As a result, the total energy of the Universe is not constrained a priori, allowing for a more general exploration of cosmological parameter space.

\begin{center}
    \begin{figure}[!t]
        \centering
    	\includegraphics[scale=0.3]{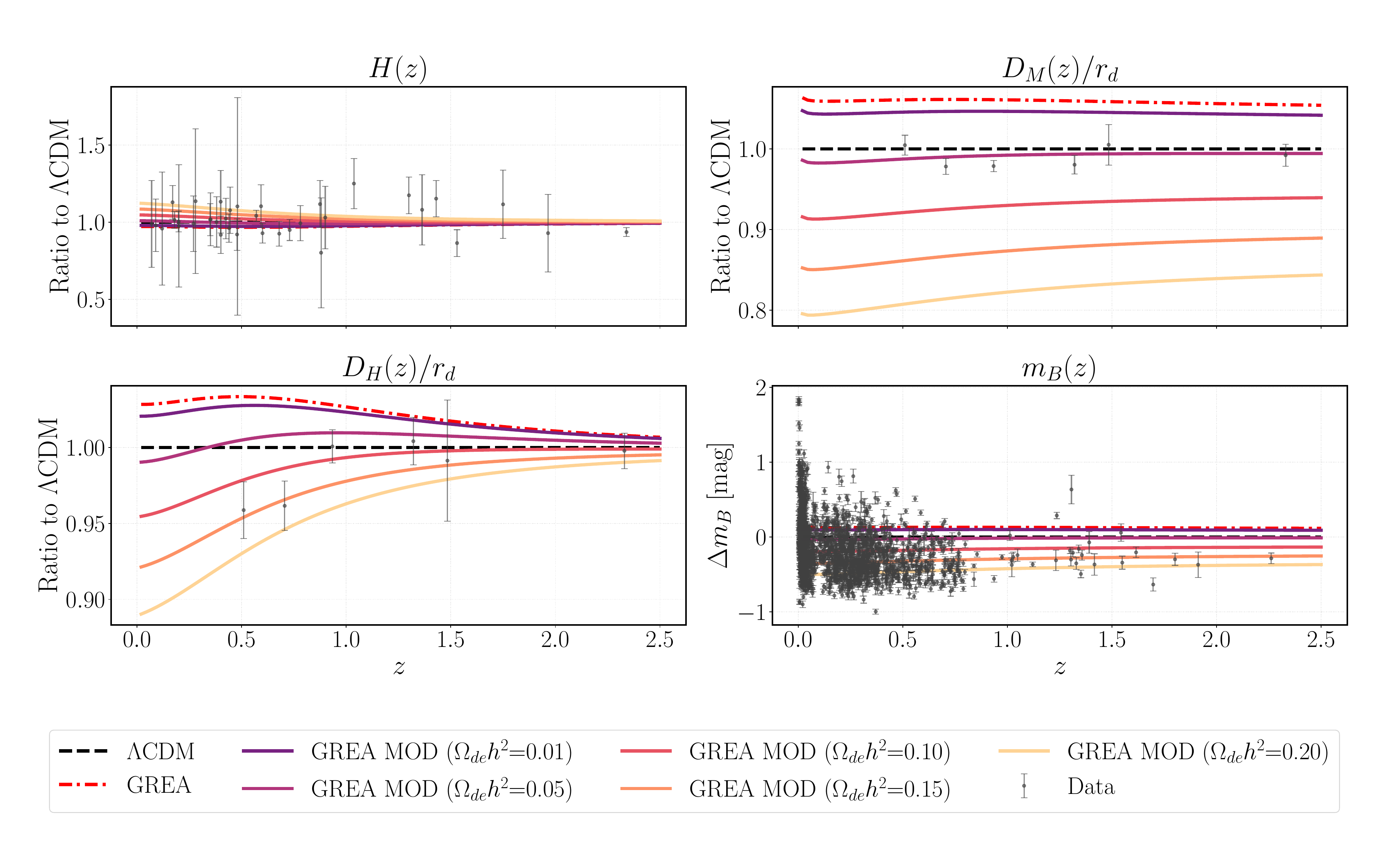}
    	\caption{\cdl{Ratios of $H(z)$ (top left), $D_M(z)/r_d$ (top right), and $D_H(z)/r_d$ (bottom left), together with the difference in apparent magnitude $\Delta m_B(z)$ (bottom right), shown for $\Lambda$CDM (black dashed), GREA (red dashed), and modified GREA (solid lines) for different values of $\Omega_{de}h^2$, all normalized to $\Lambda$CDM. Observational data are shown in each panel: cosmic chronometers (top left) \citep{Arjona_2022}, BAO (top right and bottom left) from DESI-DR2 \citep{Abdul_Karim_2025}, and Pantheon+ supernovae (bottom right) \citep{Scolnic_2022}.}
        \vspace{0.2 cm}}
        \label{fig:GREA_MOD_LAMBBDA}
    \end{figure}
\end{center}

In \autoref{fig:GREA_MOD_LAMBBDA}, we show a comparison between $\Lambda$CDM, GREA model and modified GREA for several cosmological observables, normalized to the standard scenarios.
The modification affects the late-time expansion, leading to deviations in $H(z)$, BAO distance measures, and supernova apparent magnitudes that increase with the parameter $\Omega_{de}h^2$.

We use this modified theory to test the data and to have a direct comparison between the modified GREA model, the original GREA theory, and the standard cosmological model, $\Lambda$CDM.
In particular, by fitting this extended model to both real and mock datasets, we investigate whether the data prefer a vanishing dark energy contribution, $\Omega_{de} \to 0$, indicating that GREA alone can account for the dataset, or whether $\Omega_{de}$ converges toward values compatible with those inferred within the $\Lambda$CDM model.
In this sense, the modified GREA framework interpolates between the original GREA dynamics and the standard $\Lambda$CDM, allowing for a \cdl{continous} exploration of intermediate expansion histories. 

In order to quantitatively assess the performance of these three scenarios, we also perform a Bayesian model comparison in \autoref{sec:results}.
Specifically, we compute the Bayesian evidence for each model and evaluate the corresponding Bayes factors, which allow us to determine whether the introduction of an additional dark energy component is statistically justified by the data, or whether it is disfavoured due to the increased parameter space.

\newpage
\section{Dataset}
\label{sec:dataset}

In this section, we describe in detail the dataset used in our analysis.
We first present the current observational data in \autoref{subsec:current data}, employed to constrain the model.
After that, we describe the mock datasets used for validations and forecasts in \autoref{subsec:mock_data}.

\subsection{Current observational data}
\label{subsec:current data}
For the analysis based on current data, we used four different complementary datasets, probing the expansion history of the Universe at different redshifts and with different methodologies. 
Specifically, we consider type Ia supernovae (SNIa), employed as standard candles to measure the distance of faraway objects in order to  map the expansion of the Universe; baryon acoustic oscillations (BAO), used as a standard ruler to measure angular and radial distances and to constrain the geometry and expansion history of the Universe; cosmic chronometers (CC), which provide a direct measurement of the expansion rate of the Universe, $H(z)$, at different epochs; cosmic microwave background compressed data (CMB), used to constrain the parameters of the early Universe and to provide precise calibration of the sound horizon, $r_d$.

In the following, we present the detailed description of the single current observational dataset. 

\begin{itemize}
    \item \textbf{Type Ia Supernovae (SNIa)}: 
    We used the Pantheon+ catalogue of type Ia supernovae, which consists of 1701 light curves in the redshift range $0.001<z<2.26$ \citep{Scolnic_2022}.
    SNIa are used as standard candles, providing a direct measure of apparent magnitude $m_B$, which is related to the luminosity distance $D_L(z)$ through the relation:
    \begin{equation}
    \label{eq:mb}
    m_B = 5 \text{log}_{10} \bigg( \frac{D_L(z)}{Mpc}\bigg) + 25 + M_B,
    \end{equation}
    where $M_B$ is the absolute magnitude of supernovae.
    Since $D_L(z)$ depends on the cosmology, this relation allows us to constrain the cosmological parameters.
    The luminosity distance itself is defined as: 
    \begin{equation}
    \label{eq:luminositydistance}
    D_L(z)=(1+z)D_M(z),
    \end{equation}
    where $D_M(z)$ is the comoving angular diameter distance.

    \item \textbf{Baryon Acoustic Oscillations (BAO)}: BAO measurements represent one of the most powerful distance measures in cosmology, acting as a standard ruler to map the Universe's expansion history $H(z)$, and geometric distances as a function of redshift $z$.
    Specifically, these distances are rescaled by the sound horizon at the drag epoch, $r_d$, defined as:
    \begin{equation}
    \label{eq:rd}
    r_d=\int_{z_d}^\infty \frac{c_s(z)}{H(z)}dz,
    \end{equation}
    where $c_s(z)$ is the speed of sound. 
    Before the recombination, this can be expressed as:
    \begin{equation}
    \label{eq:cs}
    c_s(z)=\frac{c}{\sqrt{3\bigg(1+\frac{3\rho_b(z)}{4\rho_{\gamma(z)}}\bigg)}},
    \end{equation}
    where $\rho_b$ and $\rho_{\gamma}$ are baryon and photon densities, respectively.
    In our analysis, the redshift of the drag epoch, $z_d$, is computed using the analytical formula developed by \citet{Aizpuru_2021}:
    \begin{equation}
    \label{eq:zdrag}
    z_{d} = \frac{1.0 + 428.169 \cdot \omega_{b}^{0.256459} \cdot \omega_{m}^{0.616388} + 925.56 \cdot \omega_{m}^{0.751615}}{\omega_{m}^{0.714129}},
    \end{equation}
    where $\omega_b= \Omega_b h^2$ and $\omega_m=\Omega_bh^2+\Omega_ch^2$.
    This approximation is justified by the fact that at high redshifts, the expansion history of the Universe for the GREA model is the same as $\Lambda$CDM.

    BAO measure the quantities $D_H/r_d$, $D_M/r_d$ and $D_V/r_d$, which are defined as:
    \begin{equation}
    \label{eq:Distances}
    D_H(z)=\frac{c}{H(z)},~~~~~  D_M(z)=\frac{c}{H_0}\int_0^z\frac{dz'}{E(z')},~~~~~D_V(z)=(zD_M(z)^2D_H(z))^{1/3}.
    \end{equation}

    In this work, we used the BAO data from DESI DR2. 
    This dataset covers seven redshift bins from $z=0.3$ to $z=2.33$, using samples from the clustering of galaxies, including Bright Galaxy Survey (BGS), Luminous Red Galaxies (LRGs), Emission Line Galaxies (ELGs), quasars (QSOs) and the Lyman-$\alpha$ forest \citep{Abdul_Karim_2025}.

    \item \textbf{Cosmic Chronometers (CC)}:
    Cosmic chronometers provide a direct and model-independent measurement of the Hubble parameter, $H(z)$.
    This method involves massive and passive galaxies, without active star formation.
    By analyzing the galaxies' spectral energy distributions, we can determine the galaxies' ages as a function of redshift \citep{Moresco_2016}.

    The fundamental principle of this method involves comparing two galaxies of this type, formed at the same time but observed at different redshifts. 
    The difference between their measured ages corresponds directly to the time elapsed between these two cosmic epochs. 
    This differential approach allows us to measure $dt/dz$, from which we can derive the Hubble parameter using the relation \citep{Moresco_2022}:
    \begin{equation}
    \label{eq:cc}
    H(z)=-\frac{1}{1+z}\frac{dz}{dt}.
    \end{equation}

    The differential nature of this method is a great advantage.  
    This makes the measurements of $H(z)$ independent of the cosmological model and less sensitive to systematic over- or underestimation of $dt$.
    All this makes cosmic chronometers an ideal tool for testing and constraining cosmological models.

    This method is not without limitations: the main challenge lies in identifying a truly passive and homogeneous sample of galaxies.
    In addition, the process of determining the age of galaxies is based on complex models of stellar evolution, and so, any uncertainty in this model, such as uncertainties in stellar metallicity or star formation history, directly affects the uncertainty in $H(z)$. 
    As a result, the statistical uncertainties in the cosmic chronometer data are greater than those obtained from other cosmological probes.

    We report the cosmic chronometer data used in our analysis in \autoref{ap:CC} \citep{Arjona_2022}.
    We assume that the measurements of $H(z)$ are independent of each other.

    \item \textbf{Cosmic Microwave Background (CMB)}:
    We used the compressed CMB information, defined by the set of parameters $\mathbf{D} = (100\,\theta_*, \omega_b, \omega_{cb})$.
    It has been found that these have nearly identical constraints on late-time dark energy as the full CMB likelihood \citep{lodha2025extendeddarkenergyanalysis,Abdul_Karim_2025}.
    Specifically, $\omega_b = \Omega_bh^2$ and $\omega_{cb} = \Omega_bh^2 + \Omega_{c}h^2$ are used to determine the sound horizon, $r_d$, which calibrates BAO measures; the angular acoustic scale $\theta_* = r(z_*)/D_M(z_*)$ provides additional geometrical information from the CMB, where $r(z_*)$ is the comoving sound horizon at the end of recombination, at redshift $z_*\approx 1089$. 
    The redshift of decoupling, $z_*$, is computed using the analytical formula developed by  \citet{Aizpuru_2021}:
    \begin{equation}
    \label{eq:z_star}
    z_*= \frac{391.672 \cdot \omega_{cb}^{-0.372296} + 937.422 \cdot \omega_b^{-0.97966}}{\omega_{cb}^{-0.0192951} \cdot \omega_b^{-0.93681}} + \omega_{cb}^{-0.731631}.
    \end{equation}
    This approximation is justified by the fact that at high redshifts, the expansion history of the Universe for the GREA model is the same of $\Lambda$CDM, because the early-universe parameters can be constrained from CMB observations almost independently of the late-time evolution \citep{Lemos_2023}. \mmt{Nevertheless, some biases on the results due to this assumption might still be present when using this information with GREA, which deviates from $\Lambda$CDM at low redshifts.}
    
    \mmt{In addition to this, the discrepancy between GREA and $\Lambda$CDM at late time necessarily yields to different predictions for $\theta^*$ between the two cosmologies. This implies that when fitting this parameter in a GREA cosmology, we will obtain values of $\omega_{cb}$ and $\omega_b$ that differ from those of $\Lambda$CDM. We will see in \autoref{subsubsec:currentdata_bayesiancomparison} how this significantly affect out results.}

    For this analysis, we used the Planck and ACT (P-ACT) combination \citep{louis2025atacamacosmologytelescopedr6}, and, in \autoref{ap:CMB}, we report the data vector and the covariance matrix \citep{calderon2025constraininggreaalternativetheory}.
\end{itemize}

\subsection{Mock datasets}
\label{subsec:mock_data}

In addition to real observations, we generate mock datasets for BAO, SNIa and gravitational waves (GW), to assess the potential constraining power of future observations on our models, and to provide a validation of the analysis pipeline.

The mock catalogue is generated using a part of the \texttt{CANDI} code  \footnote{\url{https://github.com/chiaradeleo1/CANDI}} \citep{deleo2025distinguishingdistancedualitybreaking,fazzari2025investigatingfrinflationbackgroundevolution}, which enables the construction of realistic samples that emulate expected observational features.

In particular, we generate these mock datasets using two different cosmologies with the following fiducial values:
\begin{itemize}
    \item \textbf{$\mathbf{\Lambda}$CDM mock dataset}: $H_0=69.15$ \text{km/s/Mpc} , $\Omega_bh^2=0.02253$, $\Omega_ch^2=0.1192$;
    \item \textbf{GREA mock dataset}: $H(0)=70.11$ \text{km/s/Mpc} , $\Omega_bh^2=0.02249$, $\Omega_ch^2=0.1196$, $\sqrt{-k}\eta_0=3.627$.
\end{itemize}
These parameter values are the best fit values that we obtain from the analysis of the current observational datasets BAO+SNIa+CMB+CC.
It is important to note that for the generation of the GREA based mock datasets we adopt the best fit value of the physical present day expansion rate $H(0)$, rather than the sampled parameter $H_0$.
This choice reflects the fact that, within the GREA framework, $H(0)$ is determined by the evolution of the model, as discussed in \autoref{sec:GREA_theory} and in \autoref{subsec:parameter_space}.

All the mock datasets are generated from the corresponding theoretical predictions evaluated at fiducial cosmology, while observational effects are modelled by adding a Gaussian noise to the data. For simplicity, we do not assume a correlation between the data and so, our covariance matrix can be assumed to be diagonal.

In the following we present the detailed description of all the mock datasets. 

\begin{itemize}
    \item \textbf{Type Ia Supernovae}: 
    We use mock data for SNIa, which are simulated from future observations from the Legacy Survey of Space and Time (LSST) of the Vera Rubin telescope \citep{Bianco_2021}.
    We simulate a total of $N_{SN}=8000$ events, consistent with the expected LSST number of events observed in the redshift range $z \in [0.1, 1.0]$.
    In this range we then compute the observed apparent magnitude $m_B$.
    The uncertainties are simulated by adding gaussian noise to $m_B$, with a total variance \citep{Gong_2010,Astier_2014, Hogg:2020ktc,Euclid:2020ojp,deleo2025distinguishingdistancedualitybreaking}:
    \begin{equation}
    \sigma_i^2 = \delta \mu_i^2+\sigma_{flux}^2+\sigma_{scat}^2+\sigma_{intr}^2,
    \end{equation}
    where $\sigma_{flux}=0.01$ is the contribution to the observational error due to the flux, $\sigma_{scat}=0.025$
    to the scattering and $\sigma_{intr}=0.12$ are the intrinsic uncertainties.
    
    \item \textbf{Baryon Acoustic Oscillations}: 
    We also employed mock data to investigate the constraining power of future BAO measurements, specifically those expected from the SKA Observatory (SKAO) \citep{2020}. 
    In particular, we simulate the BAO measurements assuming the optimistic survey configuration, referred to as SKA2, where the $21$-cm spectral line will be probed to extract BAO information in the redshift range $z \in [0.2, 2.0]$, divided into $18$ bins of width $\Delta z = 0.1$.
    For each redshift bin, we compute the theoretical values of BAO distances, $D_M(z)/r_d$ and $D_H(z)/r_d$, under the fiducial cosmology.
    The observational uncertainties are simulated by giving a fixed error to each quantity, based on forecasted SKA2 sensitivities.

    \item \textbf{Gravitational waves (GW)}:
    We include simulated gravitational wave (GW) standard siren observations \citep{Holz_2005, Nissanke_2010}.
    Gravitational waves from compact binary merges provide a direct measurement of the luminosity distance, independently of the cosmic ladder.
    When an electromagnetic (EM) counterpart or a host galaxy is identified, the redshift can also be measured, allowing GW events to be used as standard sirens.

    We simulate GW events from binary neutron star mergers expected from the Einstein Telescope (ET) \citep{abac2025scienceeinsteintelescope}, a third-generation ground-based detector.
    The simulations are based on a modified version of \texttt{darksirens}\footnote{\url{https://gitlab.com/matmartinelli/darksirens}} code \citep{Martinelli_2022}, assuming a monochromatic neutron star mass of $1.4\,M_{\odot}$ and a redshift-dependent merger rate, $R(z)$, consistent with population studies, defined as \citep{Cutler:2009qv}:
    \begin{equation}
        R(z) =
    \begin{cases}
    1 + 2 z, & z \leq 1,\\
    \frac{3}{4} (5 - z), & 1 < z < 5,\\
    0, & z \geq 5.
    \end{cases}
    \end{equation}
    We generate $N=20000$ injected signals and select detectable events by imposing a signal-to-noise ratio threshold $\mathrm{SNR}>8$.
    For each event, the luminosity distance and its uncertainties are computed using the public \texttt{GWFish}\footnote{\url{https://github.com/janosch314/GWFish}} code \citep{Dupletsa_2023}, employing a Fisher matrix approach.
    Regarding the identification of the EM counterpart, we follow the approach discussed in \citet{deleo2025distinguishingdistancedualitybreaking}.
    In that work, the detectability of the EM counterpart was modelled through the afterglow emission using the \texttt{afterglowpy} \footnote{\url{https://github.com/geoffryan/afterglowpy}} code \cdl{\citep{Ryan_2020}}.
    \cdl{It was shown that, in terms of the final number of events, this is equivalent to applying a cut on the inclination angle $\iota$ of the binary system, and in particular, as shown in Figure 3 of \citet{deleo2025distinguishingdistancedualitybreaking}, to reproduce the percentage of events with an associated EM counterpart it is possible to apply a cut around $\iota_{cut}=18^{\circ}$. We point out that a different choice of $\iota_{cut}$ would lead to significantly different numbers of events in our dataset, thus increasing or decreasing the constraining power of ET data.}
\end{itemize}

\subsection{Data combination}
\label{subsec:data_combination}
In our analysis, we consider different combinations of cosmological datasets, allowing us to evaluate model performance across different and complementary cosmological probes.
When using current observational data, we combine baryon acoustic oscillations (DESI), type Ia supernovae (Pantheon+), cosmic chronometers (CC) and compressed cosmic microwave background (P-ACT) measurements. 
For the forecast analyses based on mock data, we employ simulated BAO (SKAO) and SNIa (LSST) datasets, together with mock gravitational wave standard siren observations  (ET).
These combinations allow us to consistently compare the constraining power of current observations with that expected from future surveys. 
For clarity, the different dataset combinations considered in this work are summarized below:

\begin{itemize}
    \item DESI+P-ACT+Pantheon+;
    \item DESI+P-ACT+CC+Pantheon+;
    \item DESI+CC+Pantheon+;
    \item DESI+Pantheon+;
    \item  P-ACT+CC+Pantheon+;
    \item  GREA SKAO+LSST: Mock dataset generated assuming GREA as fiducial cosmology;
    \item  GREA SKAO+LSST+ET: Mock dataset generated assuming GREA as fiducial cosmology;
    \item $\Lambda$CDM SKAO+LSST: Mock dataset generated assuming $\Lambda$CDM as fiducial cosmology;
    \item $\Lambda$CDM SKAO+LSST+ET: Mock dataset generated assuming $\Lambda$CDM as fiducial cosmology.
\end{itemize}

\section{Analysis Methodology}
\label{sec:analysis}
In this section, we describe the statistical methodology adopted to constrain the parameters of the cosmological models and to perform model comparison.
Details on the likelihood construction, the numerical implementation, the sampling strategies, the prior choices and computation of Bayesian evidence are provided in the following subsections.

\subsection{Likelihood Function}
\label{subsec:likelihood}

One of the goals of our work is to estimate parameters and compare different models within a Bayesian framework.
To do this, we reconstruct the posterior probability distribution for the model parameters by evaluating the likelihood function, $\mathcal{L}$.

We assume that the cosmological datasets are statistically independent, so the total $\chi^2_{tot}$, when all the datasets are considered, is computed as the sum of the individual contributions:
\begin{equation}
\label{eq:chitot}
    \chi^2_{tot}= \chi^2_{SNIa}+ \chi^2_{BAO}+ \chi^2_{CC} + \chi^2_{CMB}+ \chi^2_{GW}.
\end{equation}
The generic $\chi^2$ for each dataset is computed as:
\begin{equation}
\label{eq:chi}
    \chi^2=\sum_{ij}(D_i-y(\vec{\theta}))C_{ij}^{-1}(D_j-y(\vec{\theta})),
\end{equation}
where $D$ is the data vector, $y(\vec{\theta})$ is the theoretical prediction dependent on the parameter vector, $\vec{\theta}$, and $C^{-1}$ is the inverse of the data covariance matrix, $C$.
The likelihood, for a Gaussian distribution, can be computed from $\chi^2$ as:
\begin{equation}
\label{eq:likelihood}
    \mathcal{L} \propto \text{exp}\bigg[-\frac{1}{2} \chi^2\bigg].
\end{equation}

\subsection{Numerical implementation and Bayesian inference}
\label{subsec:numerical_implementation}

\cdl{The theoretical implementation for the $\Lambda$CDM is handled by using the public available version of \texttt{CANDI}. While for GREA and modified GREA, we developed a custom theory module that computes the modified expansion history, this module is implemented in \texttt{CANDI}\footnote{The code for the GREA and modified GREA framework will be public upon publication.}.}

These theory modules are interfaced with the likelihoods through the  \texttt{CANDI} code, which consists of a theory module that computes background cosmological quantities for an arbitrary cosmological model, which in this work are $\Lambda$CDM, GREA and modified GREA.
In the code, a set of likelihoods for current and mock SNIa and BAO data and for future GW observations are available.
The log-likelihood evaluation is directly handled through \texttt{Cobaya} \footnote{\url{https://github.com/CobayaSampler/cobaya}} public likelihood code \citep{Torrado_2021}, interfaced with the external likelihoods.
In order to estimate the parameters of the model and derive the posterior probability distributions, we adopt a Nested Sampling approach.
This method allows for an efficient exploration of the parameter space and simultaneously provides an accurate computation of Bayesian evidence, which is essential for model comparison.
In particular, we use the \texttt{Nautilus} sampler \citep{nautilus}, configured with 4,000 live points and 16 trained neural networks, a public available code that implements the Improved Nested Sampling strategy and combines this algorithm with deep learning techniques, which is interfaced with the \texttt{Cobaya} framework to evaluate the posterior distribution and the Bayesian evidence.

\subsection{Parameters and priors}
\label{subsec:parameter_space}

\autoref{table:parameters} reports all the cosmological models considered in this work and the associated parameters, specifying for each of them whether they are sampled, derived, or fixed. 
When the parameter is sampled, we report the corresponding prior distributions. 

The sampled parameters common to all models are $H_0$, $\Omega_c h^2$, $\Omega_b h^2$.
It is important to note that although $H_0$ is sampled with a flat prior, the current Hubble expansion rate in both GREA and modified GREA is given by $H(0)$, which is obtained by solving the differential equations of the models (\autoref{eq:grea}, \autoref{eq:grea_mod}).
So, in the GREA based models, the parameter $H_0$ acts as a normalization and $H(0)$ is instead obtained by solving the dynamical equations up to $a=1$.
As a consequence, derived cosmological quantities such as $\Omega_m$, $\alpha$ and the distances are computed using $H(0)$ rather than the sampled value of $H_0$.
For consistency, we also compute $H(0)$ for the $\Lambda$CDM model, but in this case $H(0)=H_0$.
The details on the difference between $H_0$ and $H(0)$ are discussed in \autoref{sec:GREA_theory}.

\begingroup
\setlength{\tabcolsep}{10pt}
\renewcommand{\arraystretch}{1.5} 
\begin{table}[h!]
\centering
\begin{tabular}{c c c c c}
Parameter & $\Lambda$CDM & GREA & GREA MOD\\
\hline \hline
$H_0$ & $\mathcal{U}[20.0,100.0]$ & $\mathcal{U}[20.0,100.0]$ & $\mathcal{U}[20.0,100.0]$\\
$H(0)$ & derived & derived & derived\\
$\Omega_b h^2$ & $\mathcal{G}[0.02218,0.00055]$/$\mathcal{U}[0.005, 0.5]$ & $\mathcal{G}[0.02218,0.00055]$/$\mathcal{U}[0.005, 0.5]$  & $\mathcal{G}[0.02218,0.00055]$/$\mathcal{U}[0.005, 0.5]$ \\ 
$\Omega_{c}h^2$ & $\mathcal{U}[0.05, 0.5]$ & $\mathcal{U}[0.05, 0.5]$ & $\mathcal{U}[0.05, 0.5]$ \\
$\sqrt{-k}\eta_0$ & - & $\mathcal{U}[1.0, 5.0]$ & $\mathcal{U}[1.0, 5.0]$/$\mathcal{U}[1.0, 10.0]$\\
$r_{d}$ & derived & derived & derived\\
$\Omega_m$ & derived & derived & derived\\
$\Omega_{\Lambda}$ & derived & - & - \\
$\alpha$ & - & derived & derived\\
$M_B$ & $\mathcal{G}[-19.2435,0.0373]$ & $\mathcal{G}[-19.2435,0.0373]$ & $\mathcal{G}[-19.2435,0.0373]$\\
$\Omega_{de}h^2$ & - & - & $\mathcal{U}[0.0,1.0]$\\
\hline \hline
\end{tabular}
\caption{\cdl{Summary of the model parameters, their classification, and the priors adopted in the analysis. Uniform and Gaussian priors are denoted by $\mathcal{U}$ and $\mathcal{G}$.}}
\label{table:parameters}
\end{table}
\endgroup

We also used a Gaussian prior on $\Omega_b h^2$, when this quantity is externally constrained, motivated by Big Bang Nucleosynthesis (BBN) constraint \citep{Mathews_2017}.
This information is necessary when analyzing BAO data, because the BAO distance measurements mainly constrain the ratios $D_H/r_d$ and $D_M/r_d$ \citep{Adame_2025}. 
To disentangle these quantities and derive the sound horizon, $r_d$, from cosmological parameters, it is necessary to use external information, such as measurements of $\Omega_bh^2$, to break the degeneracy between \cdl{the current expansion rate} and $r_d$.
The prior on $\Omega_bh^2$ becomes flat when we include the CMB compressed information as this provides directly a measurement of this parameter.

For the GREA theory, we also used a flat prior on $\sqrt{-k}\eta_0$ and in the modified GREA, we added a flat prior on $\Omega_{de}h^2$.

A Gaussian prior on the absolute magnitude, $M_B$, should be included when analyzing supernova data.
In fact, the apparent magnitude of a supernova depends both on its absolute magnitude $M_B$ and the Hubble constant $H_0$, and the two parameters are strongly correlated.
Without additional information, it is not possible to separate the contributions of $H_0$ and $M_B$.
To break this degeneracy, we need to introduce a Gaussian prior on \cdl{the current expansion rate} or $M_B$.
In our analysis, we used a prior on $M_B$ based on the SH0ES collaboration's measurement \citep{Scolnic_2018, Riess_2022}:
\begin{equation}
\label{eq:MBprior}
    M_B= -19.2435\pm0.0373.
\end{equation}
By fixing the absolute magnitude in this way, instead of studying the Hubble tension, we focus on the supernova absolute magnitude tension \citep{Efstathiou_2021}.

The radiation density parameter is not varied in the analysis and is computed from the CMB temperature $T_{CMB}=2.7255 \texttt{K}$ \citep{Fixsen:2009ug} and assuming $N_{eff}=3.046$ \citep{Mangano:2005cc}, which is compatible with Planck results \citep{Plank_2018}.

\subsection{Bayesian evidence and model comparison}
\label{subsec:evidence}
The computation of the Bayesian evidence plays a central role in the model comparison performed in this work.
It provides a quantitative measure of how well a model is supported by the data, intrinsically accounting for both the quality of the fit and the complexity of the model itself.
The Bayesian evidence for a model $M$ and a dataset $D$ is defined as the marginal likelihood $P(D|M)$, obtained by integrating over the full parameter space  $\theta$ \citep{heavens2010statisticaltechniquescosmology}:
\begin{equation}
\label{eq:evidence}
    P(D|M)=\int P(D|\theta, M)P(\theta|M)d\theta,
\end{equation}
where $P(D|\theta, M)$ is the likelihood and $P(\theta|M)$ is the prior probability distribution of the model parameters.

A powerful method for comparing different models is the calculation of the Bayes factor, which represents the ratio between the evidence for the two competing models.
Specifically, the Bayes factor for a model $M_1$ relative to a reference model $M_2$ is defined as:
\begin{equation}
\label{eq:bayesfactor}
B_{12}=\frac{P(D|M_1)}{P(D|M_2)}.
\end{equation}
Thanks to the analysis performed with the nested sampling algorithm implemented in \texttt{Nautilus}, we were able to compute the model evidence for each model considered. 
The Bayes factor was then computed for all alternative models relative to $\Lambda$CDM, considering different data combinations, and interpreted according to the Jeffreys scale \citep{jeffreys1998theory, Robert_2009, Nesseris_2013}, summarized in \autoref{table:jeffreys}.
\begingroup 
    \setlength{\tabcolsep}{10pt}
    \renewcommand{\arraystretch}{1.5} 
    \begin{table}
        \centering
        \begin{tabular}{ c c c}  
            $B_{ij}$ & ln$B_{ij}$ & Evidence\\
            \hline \hline     
            1 $\leq$ $B_{ij}$ $\le$ 3 & 0 $\leq$ ln$B_{ij}$ $\le$ 1.1 & Weak\\
            3 $\leq$ $B_{ij}$ $\le$ 20 & 1.1 $\leq$ ln$B_{ij}$ $\le$ 3 & Definite\\
            20 $\leq$ $B_{ij}$ $\le$ 150 & 3 $\leq$ ln$B_{ij}$ $\le$ 5 & Strong\\
            150 $\leq$ $B_{ij}$ & 5 $\leq$ ln$B_{ij}$ & Very Strong\\
            \hline \hline  
        \end{tabular}  
        \caption{Jeffreys scale (\cite{Nesseris_2013}).}
        \label{table:jeffreys}
    \end{table}
\endgroup

\section{Results}
\label{sec:results}
In this section, we present the results of our analysis. 
We begin by briefly recalling the main goal of this work, namely to test the viability of the GREA framework and its extension against current and future cosmological observation.

We fist valide the GREA model using current observational data including DESI, Pantheon+, CC and P-ACT. The different data combinations adopted in the analysis are summarized in \autoref{subsec:data_combination} .
We compare our constraints with those reported in previous studies in order to assess the reliability of our implementation.
We then analyze the modified GREA model, testing the impact of an additional dark energy component.
The three framework, GREA, modified GREA and the $\Lambda$CDM, are compared using Bayesian model selection to quantitatively evaluate which model is preferred by the data.

In the second part of the analysis we focus on forecast studies based on mock dataset (see \autoref{subsec:mock_data} for details on mock data generation). 
In particular, we assume as fiducial cosmology the GREA model and we generate a mock dataset representative of next-generation-surveys, including SKAO, LSSt, and ET.
These mock data are analyzed within the GREA, modified GREA, and $\Lambda$CDM frameworks.
Finally, we consider mock datasets generated assuming $\Lambda$CDM as the fiducial cosmology, again including SKAO, LSST, and ET specifications.
We analyze these datasets within the GREA and modified GREA scenarios to assess their discriminating power.
These forecast analyses is motivated by two complementary goals. 
On the one hand, it allows us to validate the robustness and correctness of the analysis pipeline. 
On the other hand, it provides insight into how parameter estimation is affected when the data are analyzed under an incorrect cosmological model.

\subsection{Constraints from current data}
\label{subsec:result current data}
We begin by validating our implementation of the GREA framework against current observational datasets in order to verify the numerical stability of the code and the consistency of our results with previous analyses in the literature.
This step is essential to ensure that our numerical implementation correctly reproduces the phenomenology of the model.

By fitting the GREA theory to the DESI+P-ACT+Pantheon+ combination, we obtain parameter constraints that are in good agreement with previous studies.
In particular, the inferred values of the GREA parameters, reported in \autoref{table:results}, are consistent with those reported in the literature \citep{calderon2025constraininggreaalternativetheory} within $1\sigma$, confirming the reliability of our analysis pipeline.

In addition to this, we extend the analysis to the DESI+P-ACT+CC+Pantheon+, which also includes the cosmic chronometers.
The corresponding best fit values are reported in \autoref{table:results}.

\begin{center}
    \begin{figure}[!t]
        \centering
    	\includegraphics[scale=0.3]{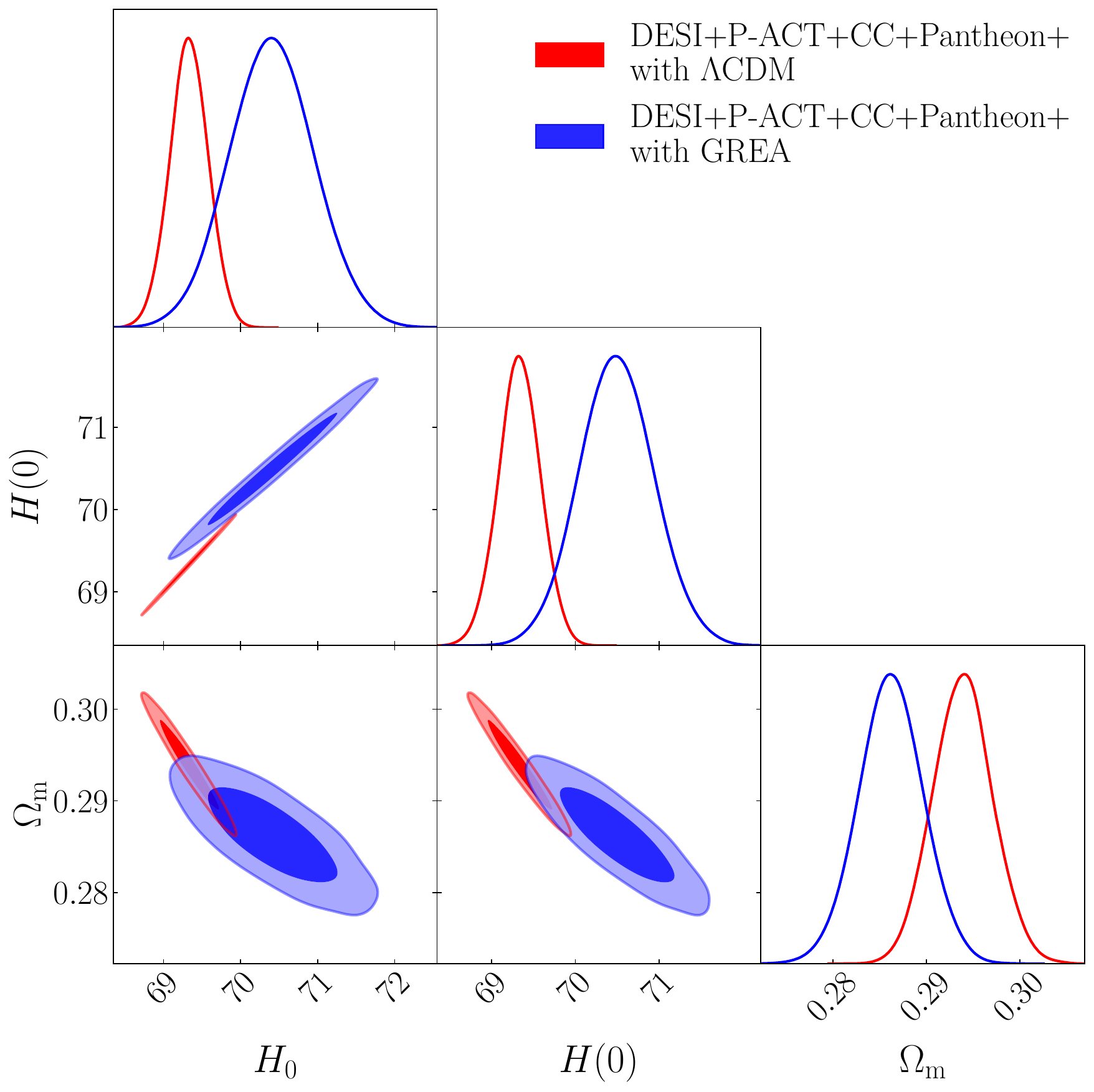}
        \includegraphics[scale=0.3]{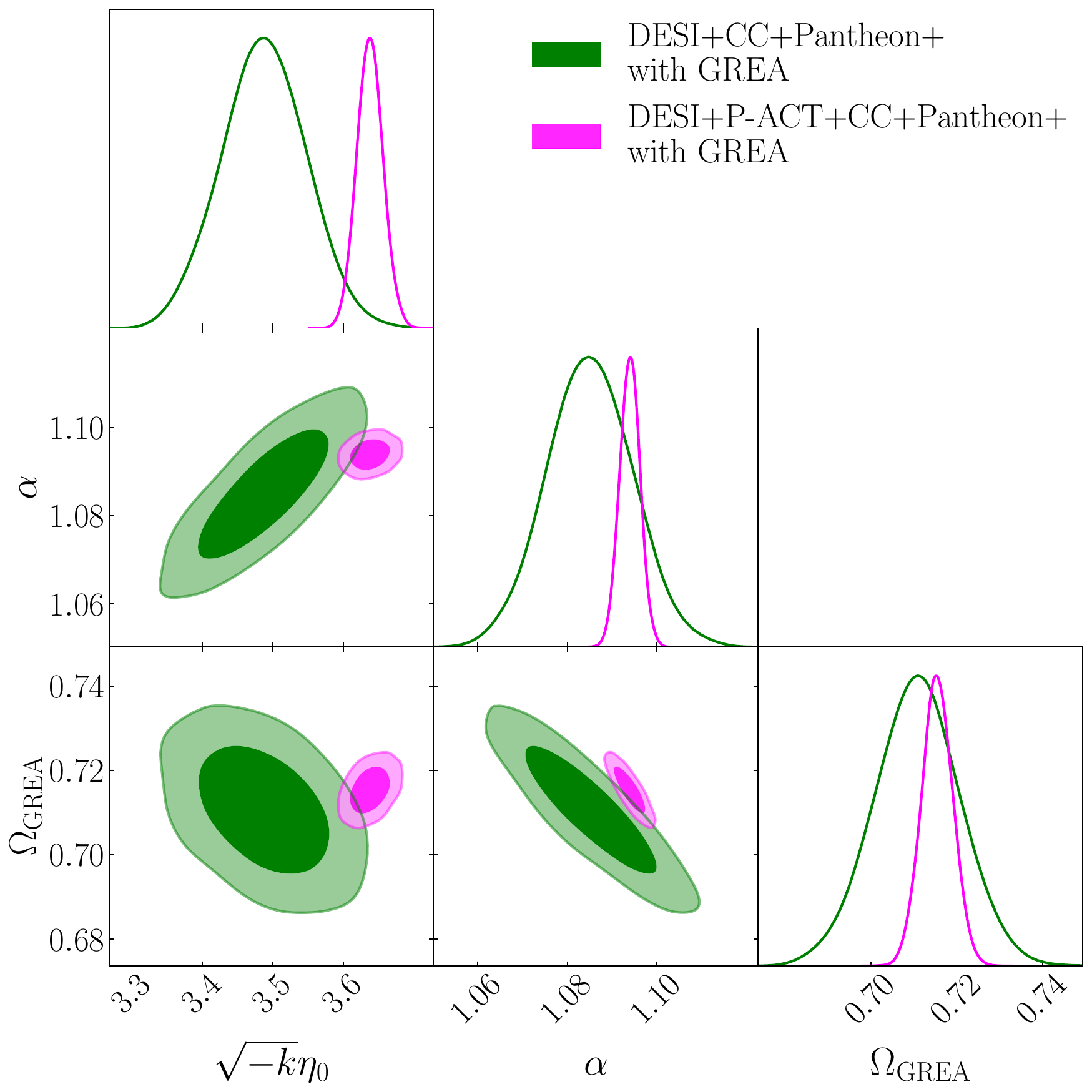}
    	\caption{Left panel: posterior distribution of $H_0$, $H(0)$ and $\Omega_m$ from DESI+P-ACT+CC+Pantheon+ obtained with $\Lambda$CDM (red) and GREA (blue).
        Right panel: posterior distribution of $\sqrt{-k}\eta_0$, $\alpha$ and $\Omega_{GREA}$ from DESI+P-ACT+CC+Pantheon+ (pink) and DESI+CC+Pantheon+ (green) obtained with GREA theory. 
        \vspace{0.2 cm}}
        \label{fig:figura_1}
    \end{figure}
\end{center}

We explicitly compare the GREA results with the standard $\Lambda$CDM model, highlighting the differences in the inferred cosmological parameters.
In particular, in \autoref{fig:figura_1} (left panel) we present the posterior distribution of $H_0$, $H(0)$ and $\Omega_m$ for both models.
We find that GREA prefers a present-day expansion rate that is systematically larger than the value inferred within $\Lambda$CDM for the same dataset.
This shift pushes the predicted \mmt{current expansion rate} toward higher values, potentially shifting the inferred expansion rate toward values closer to those measured by local probes. \mmt{Nevertheless, this shift towards higher rates of expansion is only able to ease the tension with local measurements, but the tension remains statistically significant.}

Furthermore, the two models exhibit distinct degeneracy directions in the $H(0)-\Omega_m$ plane; specifically, GREA allows for a broader range of \mmt{values for the current expansion rate} while favoring lower values of $\Omega_m$ compared to $\Lambda$CDM.

In \autoref{fig:figura_1} (right panel) we also show the $(\alpha, \sqrt{-k}\eta_0, \Omega_{GREA})$ parameter space for DESI+P-ACT+CC+Pantheon+ and DESI+CC+Pantheon+.
The GREA parameters, $\alpha$ and $\sqrt{-k}\eta_0$, exhibit a pronounced degeneracy, particularly in the absence of CMB compressed information. 
In the following analysis, we therefore adopt $\sqrt{-k}\eta_0$ as the primary geometric parameter characterizing the GREA framework.
We can also notice that the value of $\Omega_{GREA}$ is very similar to the value of $\Omega_{\Lambda}$ in the standard $\Lambda$CDM model \mmt{across all data combinations} (\autoref{table:results}), indicating that the entropic force effectively mimics the role of a cosmological constant in the late-time expansion history. \mmt{This highlights how, despite the different expansion at low redshift, GREA still produces a redshift trend of the energy densities which is consistent with that of the standard model.}

We now turn to the analysis of the modified GREA framework, in which an additional dark energy component parametrized by $\Omega_{de}$ is introduced.
In particular, in our analysis we let $\Omega_{de}h^2$ free with a flat prior.

\begin{center}
    \begin{figure}[!t]
        \centering
        \includegraphics[scale=0.5]{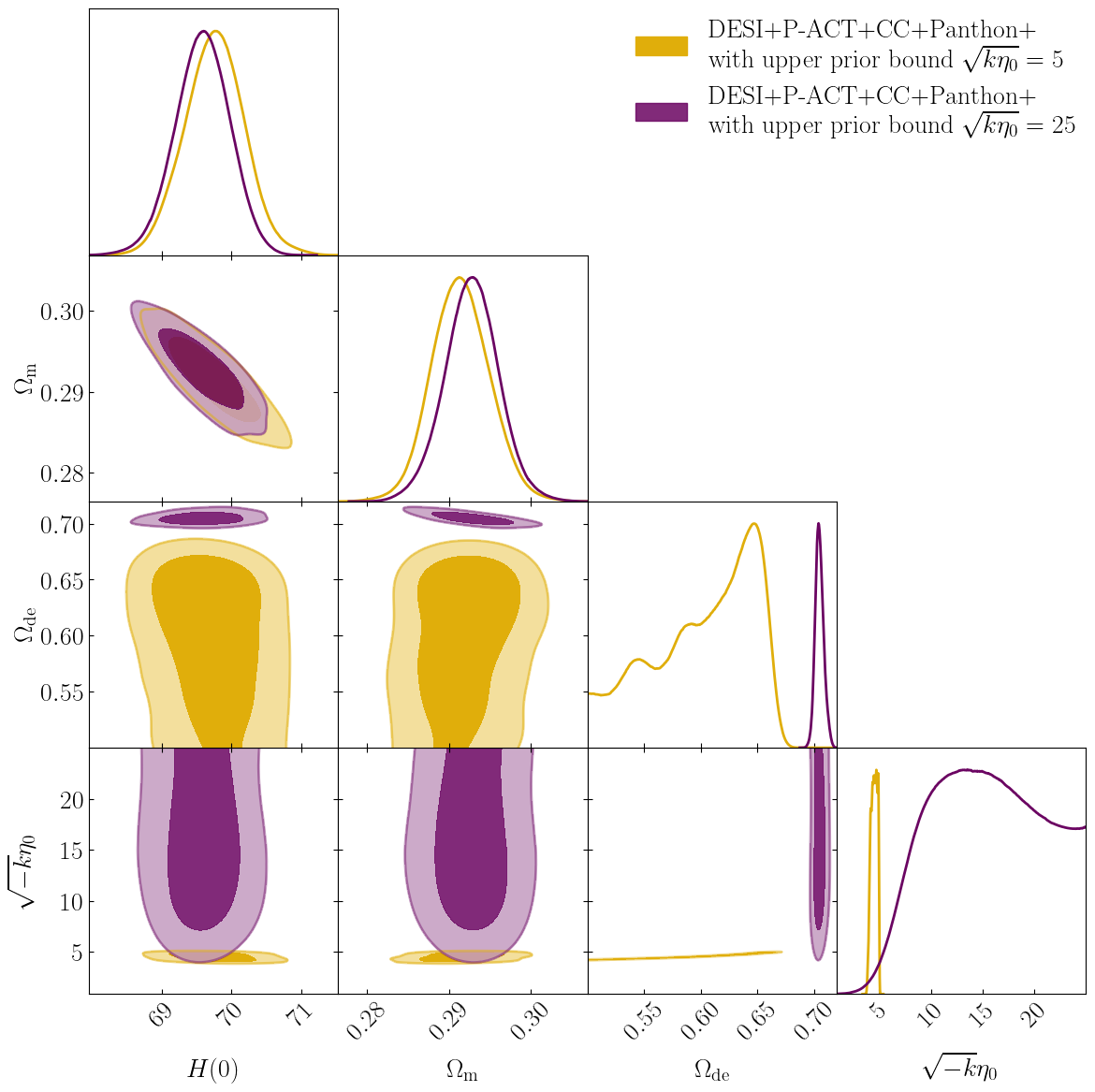}
    	\caption{Posterior distribution of $H(0)$, $\Omega_m$ and $\Omega_{de}$ from DESI+P-ACT+CC+Pantheon+ obtained with modified GREA. 
        In the orange case we use a flat prior on $\sqrt{-k}\eta_0$ between $[1.0;5.0]$, in the purple case we use a flat prior on $\sqrt{-k}\eta_0$ between $[1.0;25.0]$, allowing for the $\Lambda$CDM limit.
        \vspace{0.4 cm}}
        \label{fig:figura_2}
    \end{figure}
\end{center}

The posterior distributions for DESI+P-ACT+CC+Pantheon+ are shown in \autoref{fig:figura_2} in orange.
The extra dark energy parameter, $\Omega_{de}$, exhibits a highly asymmetric posterior distribution, with a pronounced degeneracy with the GREA parameters, in particular with $\sqrt{-k}\eta_0$.
Moreover, the characterizing parameter of GREA, $\sqrt{-k}\eta_0$, is very weakly constrained.

\begin{center}
    \begin{figure}[!t]
        \centering
        \includegraphics[scale=0.8]{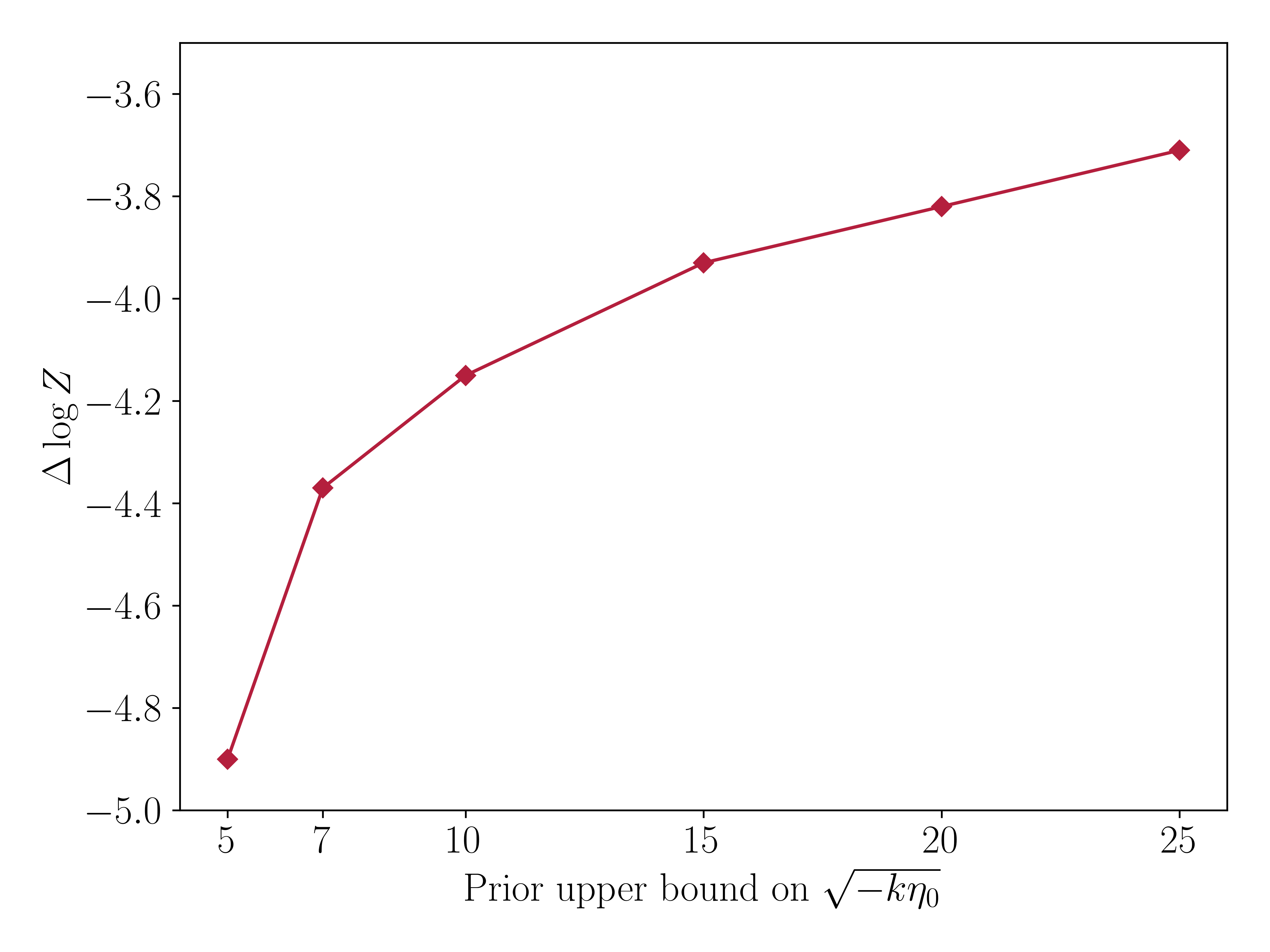}
    	\caption{\cdl{Impact of the upper limit of the prior range for $\sqrt{-k\eta_0}$ using the full combined dataset (DESI+P-ACT+CC+Pantheon) within the modified GREA model. The $\Delta \log Z$ is computed relative to the same data combination within the $\Lambda$CDM model.}
        \vspace{0.4 cm}}
        \label{fig:logZ_upper_limit}
    \end{figure}
\end{center}

To better understand the origin of these degeneracies, we investigate the use of a large prior on $\sqrt{-k}\eta_0$, extending its upper bound from $5$ to \cdl{$25$}. 
\cdl{In particular, as }shown in figure \autoref{fig:figura_2} in purple, enlarging the prior volume significantly affects the posterior distribution of $\Omega_{de}$, while leaving the constraints on the standard cosmological parameters such as $H(0)$ and $\Omega_m$ largely unchanged.
Enlarging the prior volume allows the analysis to explore regions in which the entropic contribution becomes subdominant, effectively approaching the $\Lambda$CDM limit of the modified GREA framework.
\cdl{We conduct a preliminary analysis to assess the impact of the upper bound of the prior by examining the evidence across the combined datasets (DESI+P-ACT+CC+Pantheon) in both the $\Lambda$CDM and modified GREA models. Specifically, we run the modified GREA model using different upper bounds for $\sqrt{-k\eta_0}$: [5, 7, 10, 15, 20, 25]. We know that when the data points towards the $\Lambda$CDM model, as shown in \autoref{table:evidence_realdata}, the corresponding limit for $\sqrt{-k\eta_0}$ would be infinite. As a result, if the analyze data have a preference for $\Lambda$CDM over GREA, the contours would not close and would instead point toward higher values of $\sqrt{-k\eta_0}$. However, as demonstrated in \autoref{fig:logZ_upper_limit}, we observe that $\Delta \log Z$ decreases and eventually stabilizes as the upper bound increases, which aligns with expectations from $\Lambda$CDM-favored data.
Given this understanding, for the analysis of the modified GREA model, we adopt an upper bound of $\sqrt{-k\eta_0} = 25$.}
The best fit values are reported in \autoref{table:results}.

\newpage
\subsubsection{Bayesian model comparison}
\label{subsubsec:currentdata_bayesiancomparison}
In addition to parameter estimation, we perform a Bayesian model comparison among the three competing frameworks: $\Lambda$CDM, GREA and modified GREA.
The Bayesian evidence for each model is computed as described in \autoref{subsec:evidence}, and Bayes factor is evaluated relative to $\Lambda$CDM reference model as 
\begin{equation}
    \Delta \text{log} Z = \text{log} Z_{model}-\text{log} Z_{\Lambda CDM}
\end{equation}

Our results, for all the current data combinations, are reported in \autoref{table:evidence_realdata}.
These combinations allow us to isolate the impact of different observational probes, in particular the role of CMB compressed data, on the statistical performance of the model.
 
\begin{center}
    \begin{figure}[!t]
        \centering
         \includegraphics[width=0.8\linewidth]{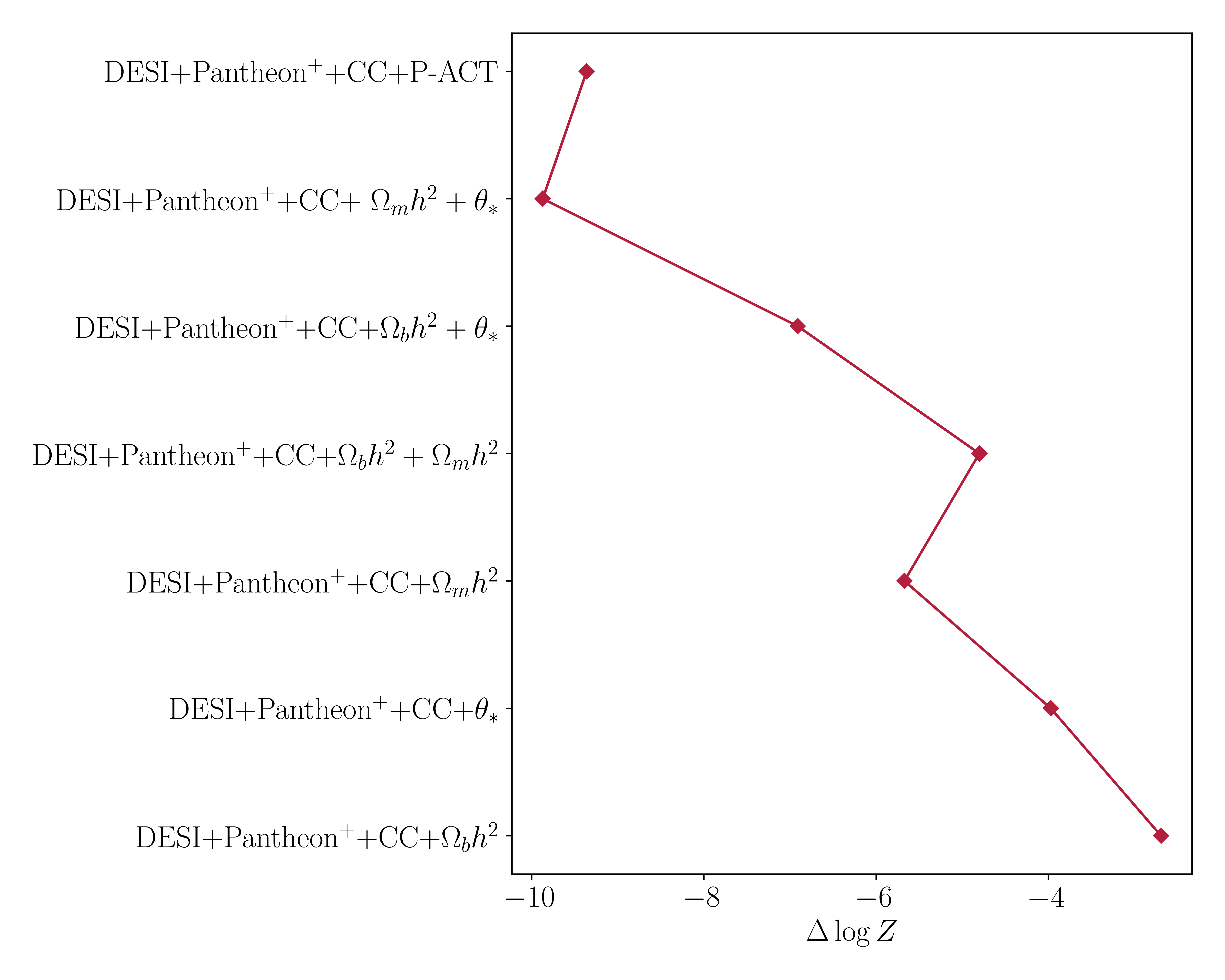}
    	\caption{\cdl{$\Delta \log Z$ evaluated for the GREA model with DESI+CC+Pantheon+ data, adding one parameter of the CMB P-ACT ($\theta_*$, $\Omega_B h^2$ and $\Omega_m h^2$) at a time, then two parameters, and finally the full CMB dataset. The $\Delta \log Z$ is evaluated with respect to the corresponding combination of data analyzed in $\Lambda$CDM model.}
        \vspace{0.4 cm}}
        \label{fig:logZ_CMB}
    \end{figure}
\end{center}

\cdl{With respect to the DESI+CC+Pantheon+ dataset, the inclusion of CMB data in the GREA analysis strongly disfavors the GREA model, with $\Delta \log Z = -9.460$ compared to $\Delta \log Z = -2.764$ when CMB data is excluded. As explained above, the CMB data provides compressed information from P-ACT on $100\theta_*$, $\omega_b$, and $\omega_{cb}$. We now investigate which parameters are driving the data to strongly favor the $\Lambda$CDM model. Specifically, we perform the GREA analysis using the DESI+CC+Pantheon+ dataset and sequentially add CMB information, starting with one parameter at a time, then two parameters, and finally the full CMB dataset. The results on $\Delta \log Z$ are presented in \autoref{fig:logZ_CMB}, where the evidence is calculated relative to the corresponding combination of data analyzed with the $\Lambda$CDM model. We observe that no single parameter alone strongly disfavours the GREA model in comparison to the $\Lambda$CDM case. However, when the information on $\theta_*$ is combined with $\Omega_b h^2$ or $\Omega_m h^2$, the GREA model is strongly disfavored. This can be attributed to the fact that $\theta_*$ is computed from the angular diameter distance over a redshift range typically spanning from the early universe at $z \sim 1000$, where the GREA model does not introduce any modifications to the Hubble expansion, to the late universe at $z = 0$, where the Hubble expansion is modified according to the GREA model. During this period, the modified Hubble expansion in the GREA model leads to differences in the inferred values of $\theta_*$. On the other hand, the inclusion of information about the matter content, such as $\Omega_b h^2$ and $\Omega_m h^2$, which are not modified in the GREA model and are consistent with the $\Lambda$CDM model, disfavors the GREA model when combined with $\theta_*$. }
The modified GREA model shows a significant improvement with respect to the original GREA, with $\Delta \text{log} Z = -3.71$, corresponding to a less significant evidence for $\Lambda$CDM.
This behavior is due to the fact that modified GREA allows $\Lambda$CDM as a limit but has two additional parameters relative to the standard model.

To further investigate the origin of this preference for $\Lambda$CDM, we also test the models using data combinations that do not include CMB observations.
For DESI+CC+Pantheon+ and DESI+Pantheon+, the evidence against GREA is reduced, with $\Delta \text{log} Z = -2.76$ and $-1.85$, respectively.
This result indicates that CMB information plays a dominant role in driving the strong statistical preference for $\Lambda$CDM.
A similar trend is observed for the modified GREA model from DESI+CC+Pantheon+, whose evidence relative to $\Lambda$CDM becomes weaker when CMB data are excluded.

For the P-ACT+CC+Pantheon+ dataset, which includes CMB information but excludes BAO, both GREA and modified GREA are disfavoured, with $\Delta \text{log} Z = -15.36$ and $-1.69$, respectively. \cdl{In particular, the result obtained for GREA is even more negative than the one found for DESI+P-ACT+Pantheon+, for which $\Delta \text{log} Z = -9.46$. This suggests that, once BAO information is removed, the CMB angular scale $\theta_\ast$ becomes the dominant geometrical constraint. Since $\theta_\ast$ is sensitive to both the matter density and the integrated expansion history up to recombination, GREA has difficulty finding values of $\Omega_m$ and $\sqrt{-k}\eta_0$ that simultaneously reproduce the CMB distance information and remain compatible with the low-redshift CC and Pantheon+ data. This tension leads to a stronger statistical preference for $\Lambda$CDM.}

However, these results suggest that the inclusion of the extra dark energy component partially alleviates the tension with between observations, but it is not sufficient to fully compete with $\Lambda$CDM when CMB compressed data are included.
It is important to emphasize that these results are obtained using compressed CMB information, which provides a robust but simplified representation of the high redshift constraints, but it does not represent the complete information contained in the full CMB power spectrum.

Overall, current background observations do not provide statistical support for GREA or its modified extension over $\Lambda$CDM. 
However, the reduced level of disfavor in the absence of CMB information suggests that future high-precision low-redshift probes may provide additional discriminatory power, and that a more complete analysis of CMB data, using the full likelihood and testing the evolution of scalar perturbations in GREA, is required.

\begingroup 
    \setlength{\tabcolsep}{10pt}
    \renewcommand{\arraystretch}{1.5} 
    \begin{table}
        \centering
        \begin{tabular}{ c c c }  
            Model & Data combination & $\Delta \text{log} Z$\\
            \hline \hline  
            $\Lambda$CDM & DESI+P-ACT+Pantheon+ & 0 \\
            GREA & DESI+P-ACT+Pantheon+ & -9.460\\
            Modified GREA & DESI+P-ACT+Pantheon+ & -3.648\\
            $\Lambda$CDM & DESI+P-ACT+CC+Pantheon+ & 0 \\
            GREA & DESI+P-ACT+CC+Pantheon+ & -9.364\\
            Modified GREA & DESI+P-ACT+CC+Pantheon+ & \cdl{-3.71}\\
            $\Lambda$CDM & DESI+CC+Pantheon+ & 0 \\
            GREA & DESI+CC+Pantheon+ & -2.764\\
            Modified GREA & DESI+CC+Pantheon+ & \cdl{-1.85}\\
            $\Lambda$CDM & DESI+Pantheon+ & 0 \\
            GREA & DESI+Pantheon+ & -1.721\\
            $\Lambda$CDM & P-ACT+CC+Pantheon+ & 0 \\
            GREA & P-ACT+CC+Pantheon+ & -15.365\\
            Modified GREA & P-ACT+CC+Pantheon+ &\cdl{-1.69}\\
            \hline \hline  
        \end{tabular}  
        \caption{Bayesian evidence and Bayes factor for $\Lambda$CDM, GREA and modified GREA obtained from the analysis of all the current observational data combinations. The Bayes factor is computed respect to $\Lambda$CDM as $\Delta \text{log} Z= \text{log}Z_{model}-\text{log}Z_{\Lambda CDM}$.}
        \label{table:evidence_realdata}
    \end{table}
\endgroup

\subsection{Forecasts with GREA-based mock data}
\label{subsec:mock_GREA}

To evaluate the constraining power of future observations and to test the internal consistency of the GREA framework, we perform a forecast analysis using mock datasets generated assuming GREA as the fiducial cosmological model (GREA SKAO+LSST, GREA SKAO+LSST+ET).
In particular, we focus on mock gravitational wave standard siren data, which directly probe the luminosity distance-redshift relation. The mock datasets are analyzed using the GREA theory, its modified extension, and $\Lambda$CDM.

\subsubsection{Recovery of fiducial parameters}
\label{subsubsec:GREAmock_recovery}
\begin{center}
    \begin{figure}[!t]
        \centering
    	\includegraphics[scale=0.3]{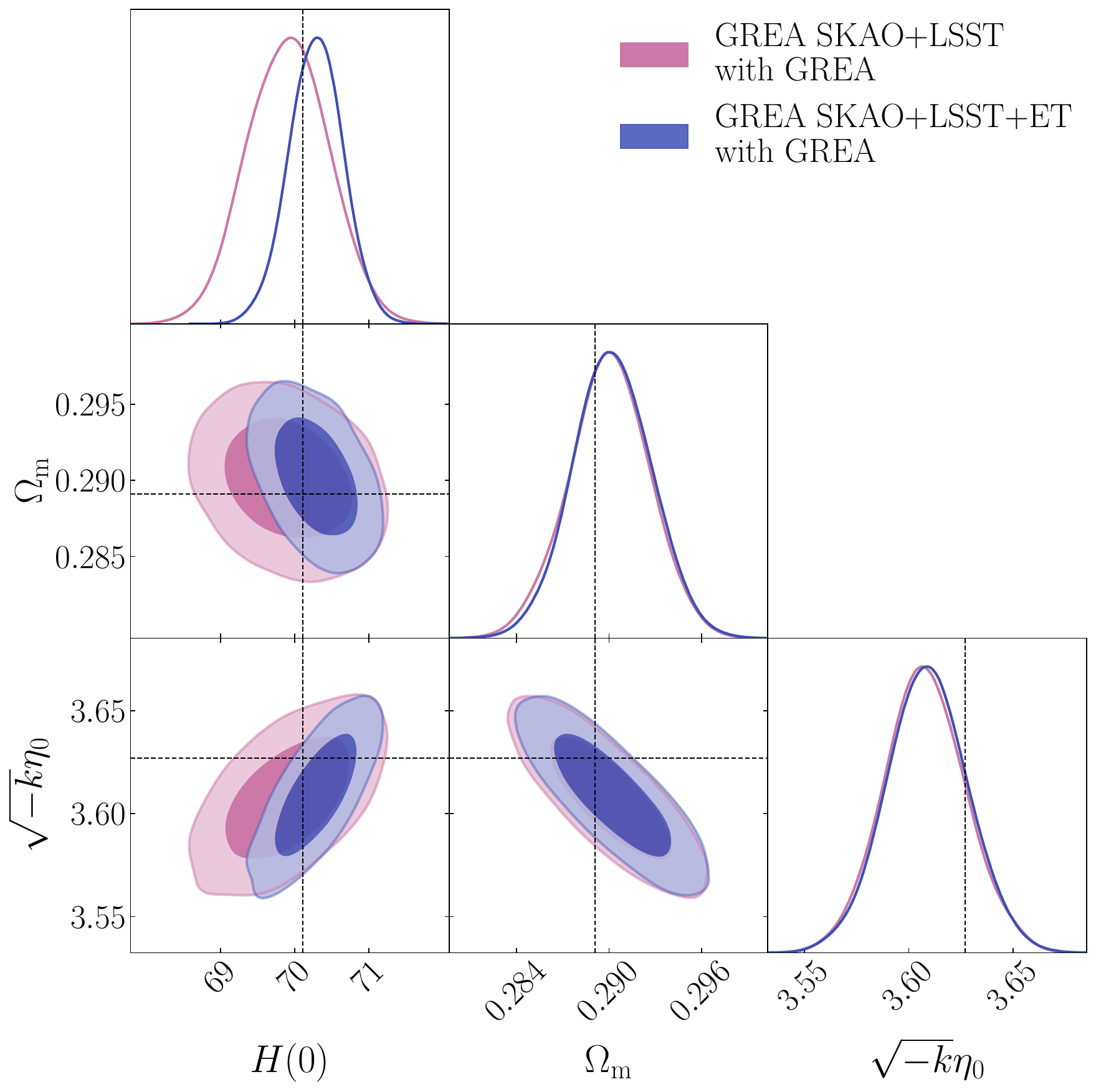}
        \includegraphics[scale=0.3]{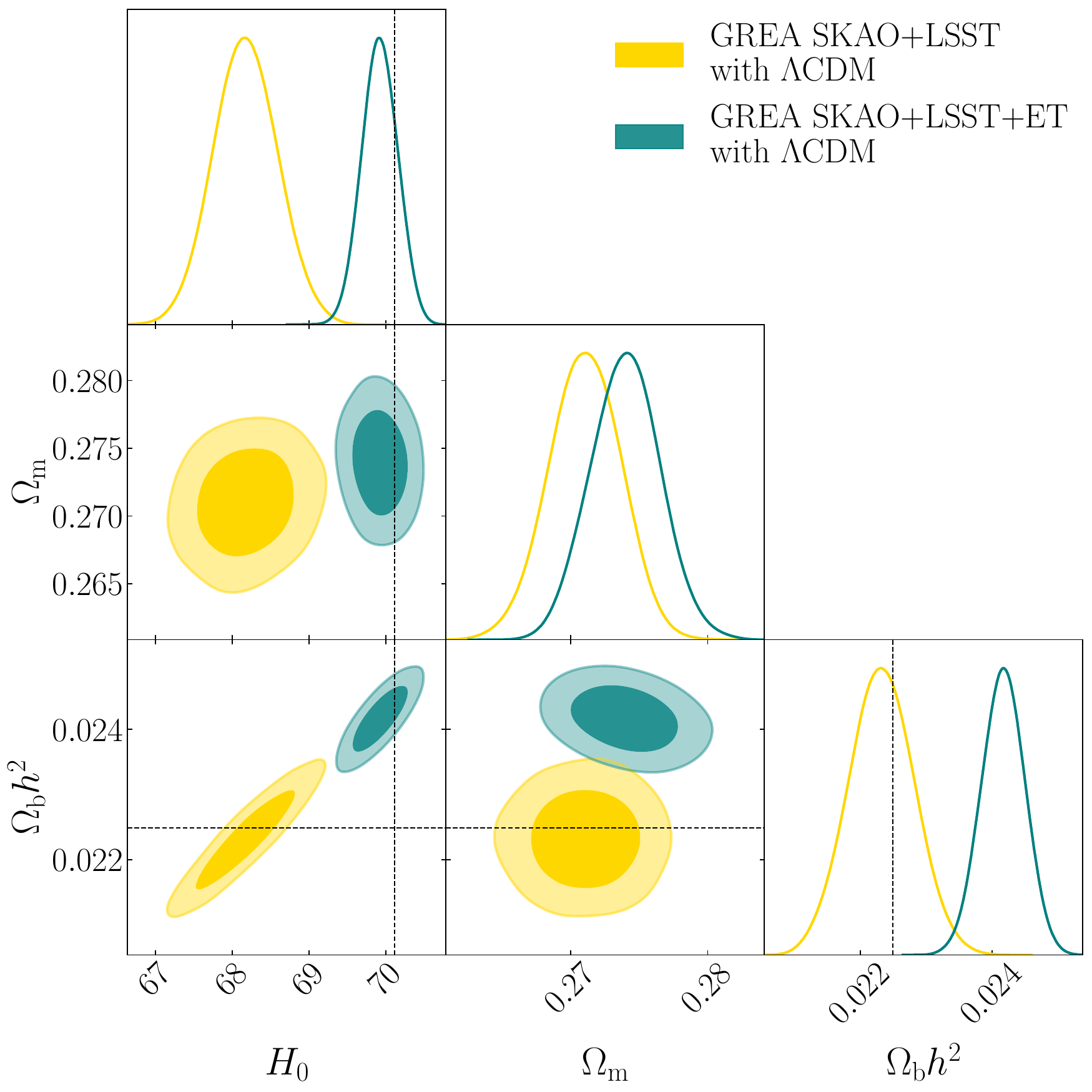}
    	\caption{Left: posterior distribution of GREA parameters from GREA SKAO+LSST (pink), GREA SKAO+LSST+ET (blue) obtained with GREA.
        The dashed lines correspond to the parameters value that we used to generate the mock datasets.
        Right: posterior distribution of $\Lambda$CDM parameter from GREA SKAO+LSST (yellow), GREA SKAO+LSST+ET (green) obtained with $\Lambda$CDM analysis.
        The dashed lines correspond to the parameters values that we used to generate the mock datasets.
        \vspace{0.3 cm}
        }
        \label{fig:figura_4}
    \end{figure}
\end{center}

We first analyze the GREA-based mock datasets assuming the correct underlying model, i.e. GREA.
This consistency test verifies that the inference pipeline is able to recover the input fiducial parameters without introducing spurious biases.
As shown in \autoref{fig:figura_4} (left panel), the input fiducial parameters are accurately recovered for both data combinations.
The dashed lines indicate the values used to generate the mocks, and they lie well within the posterior contours. 
The inclusion of GW data (SKAO+LSST+ET) significantly tightens the constraints, reducing parameter degeneracies and improving overall precision.

We then analyze the same mock datasets assuming $\Lambda$CDM cosmology.
In this case, \autoref{fig:figura_4} (right panel), the recovered parameters exhibit systematic shifts with respect to the GREA fiducial values.
In particular, when we include GW data, the posterior distribution shows a preference for larger values of $H(0)$.
This shift reflects the systematic difference between the luminosity distances predicted by GREA and those predicted by $\Lambda$CDM.
Since GW standard sirens provide a direct measurement of $D_L(z)$, they are especially sensitive to such discrepancies.

This comparison demonstrates that if the true cosmology is described by GREA, future datasets, and especially GW observations, would reveal statistically significant tensions when analyzed under an incorrect $\Lambda$CDM assumption.

\subsubsection{Impact of GW data in the modified GREA framework}
\label{subsubsec:GREAmock_GW}
We now analyze the GREA-based mock dataset GREA SKAO+LSST+ET within the modified GREA framework, which introduces an additional dark energy component parametrized by $\Omega_{de}$.

Since the fiducial cosmology does not contain an explicit dark energy term, this test allows us to assess whether future data will be able to rule out a cosmological constant contribution.

\begin{center}
    \begin{figure}[!t]
        \centering
        \includegraphics[scale=0.28]{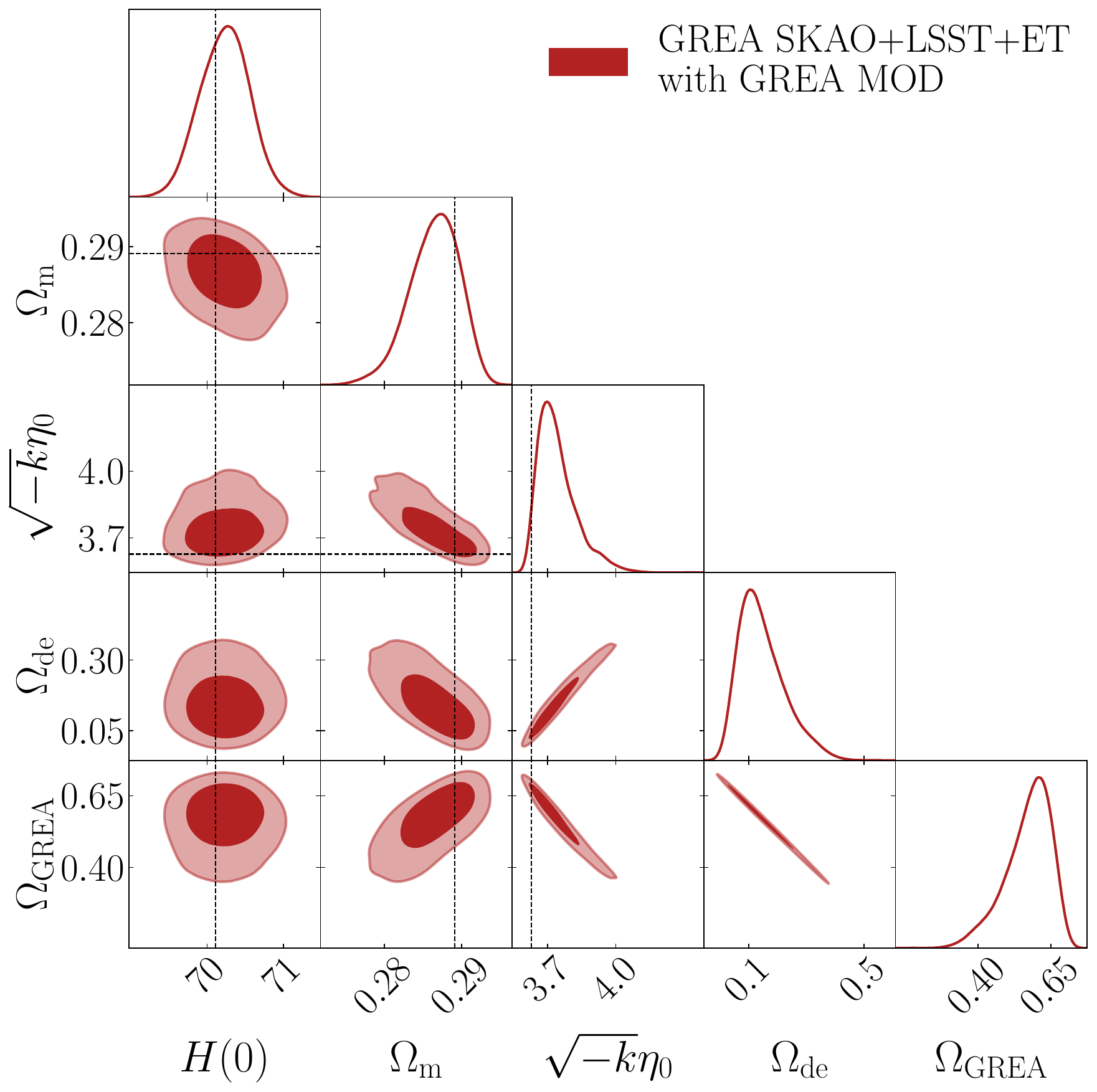}
        \includegraphics[scale=0.23]{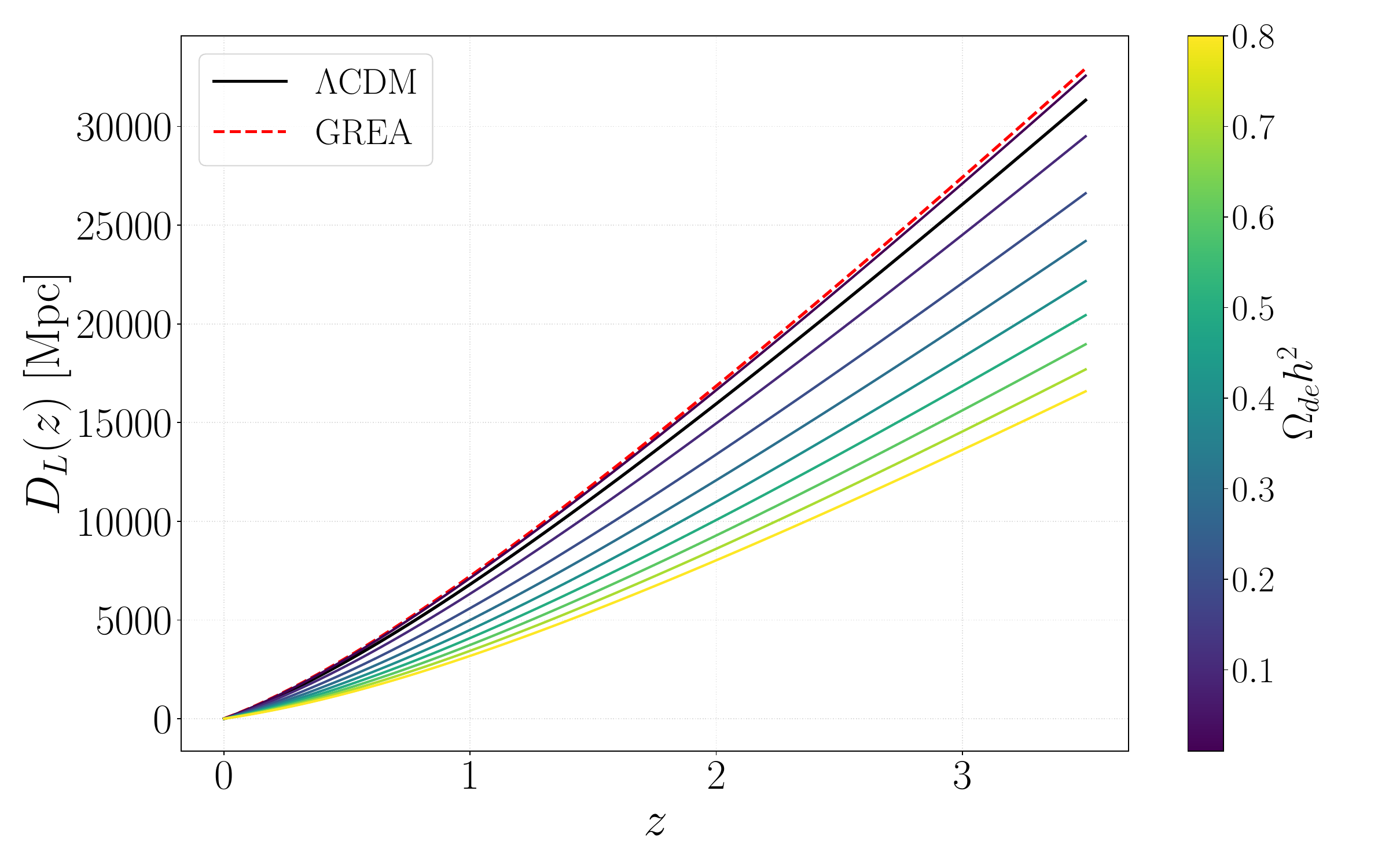}
    	\caption{Left panel: Posterior distribution of modified GREA parameter from GREA SKAO+LSST+ET.
        The dashed lines correspond to the parameters value that we used to generate the mock datasets.
        Right panel: Luminosity distance, $D_L(z)$ predicted by $\Lambda$CDM (black line), GREA (dashed red line) and modified GREA for different values of $\Omega_{de}h^2$.
        \vspace{0.2cm}
        }
        \label{fig:figura_5}
    \end{figure}
\end{center}

Our results are shown in \autoref{fig:figura_5} (left panel) and reported in \autoref{table:results}.
In particular, we obtain $\Omega_{de}=0.146^{+0.051}_{-0.098}$ and $\Omega_{GREA}=0.571^{+0.096}_{-0.051}$.
The posterior distribution of $\Omega_{de}$ is asymmetric toward low values and remains compatible with $\Omega_{de}=0$ within $2\sigma$.
This indicates that, when the true cosmology is described by GREA, the modified theory does not require an additional dark energy component once GW information is included.
The inclusion of GW also allows us to correctly recover the best fit of GREA parameters (dashed lines).
For instance, $H(0)=70.23\pm 0.33$ and $\Omega_m= 0.2865^{+0.0036}_{-0.0028}$, showing tight and symmetric constraints.
This reflects the strong sensitivity of GW standard sirens to the luminosity distance-redshift relation.
The physical origin of this behavior can be understood by examining the luminosity distance $D_L(z)$ shown in \autoref{fig:figura_5} (right panel).
In the modified GREA, a non-zero $\Omega_{de}$ can partially mimic the expansion history induced by the entropic component.
However, GW observations provide absolute measurements of $D_L(z)$, directly constraining the expansion rate and efficiently suppressing deviations from the fiducial GREA predictions.

\subsubsection{Bayesian model comparison}
\label{subsubsec:GREAmock_bayesiancomparison}
The discriminating power of future observations can be quantified through a Bayesian model comparison, which allows us to assess whether the additional dark energy component is statistically justified once that model complexity is taken in account.

The results are summarized in \autoref{table:evidence_mockdataGREA}.
We compute the Bayesian evidence with respect to GREA as:
\begin{equation}
    \Delta \text{log} Z = \text{log} Z_{model}-\text{log} Z_{GREA}
\end{equation}
For the GREA SKAO+LSST mock dataset analyzed under $\Lambda$CDM we obtain $\Delta \text{log} Z = -3.59$, indicating a strong statistical preference for GREA over $\Lambda$CDM according to Jeffreys scale.

For the GREA SKAO+LSST+ET combination, which includes GW, the discriminating power increases.
In this case, $\Lambda$CDM is strongly disfavoured with respect to GREA, with  $\Delta \text{log} Z = -7.95$, while the modified GREA is moderately disfavoured with $\Delta \text{log} Z = -5.65$.
In these cases, the high values of $\Delta \text{log} Z$ are also justified by the behavior of $D_L(z)$ (\autoref{fig:figura_5}, right panel).
When the fiducial cosmology is described by GREA, the inclusion of high-precision GW observations strengthens the statistical preference for the correct underlying model and penalizes alternative descriptions that introduce unnecessary degrees of freedom.

Overall, this forecast analysis demonstrates that future multi-probe observations, and in particular GW standard sirens, will provide sufficient sensitivity to correctly identify GREA as the preferred model when it represents the true cosmological scenario.

\begingroup 
    \setlength{\tabcolsep}{10pt}
    \renewcommand{\arraystretch}{1.5} 
    \begin{table}
        \centering
        \begin{tabular}{ c c c c }  
            Model & Data combination  & $\Delta \text{log} Z$\\
            \hline \hline  
            $\Lambda$CDM & GREA SKAO+LSST & -3.594\\
            GREA & GREA SKAO+LSST &  0\\
            $\Lambda$CDM & GREA SKAO+LSST+ET & -7.945\\
            GREA & GREA SKAO+LSST+ET & 0\\
            Modified GREA & GREA SKAO+LSST+ET  & \cdl{-5.650} \\ 
            \hline \hline  
        \end{tabular}  
        \caption{Bayesian evidence and Bayes factor for $\Lambda$CDM, GREA and modified GREA obtained from the analysis of GREA SKAO+LSST and GREA SKAO+LSST+ET. The Bayes factor is computed respect to GREA as $\Delta \text{log} Z= \text{log}Z_{model}-\text{log}Z_{GREA}$.}
        \label{table:evidence_mockdataGREA}
    \end{table}
\endgroup

\subsection{Testing GREA models with $\Lambda$CDM-based mock data}
\label{subsec:mock_LCDM}

We generate mock datasets assuming $\Lambda$CDM as the fiducial cosmology ($\Lambda$CDM SKAO+LSST, $\Lambda$CDM SKAO+LSST+ ET) and analyze them within the GREA and modified GREA frameworks.

This analysis allows us to assess whether the GREA and modified GREA frameworks can reproduce a $\Lambda$CDM-like expansion history and to what extent future observations can statistically distinguish between these scenarios.

\begin{center}
    \begin{figure}[!t]
        \centering
    	\includegraphics[scale=0.5]{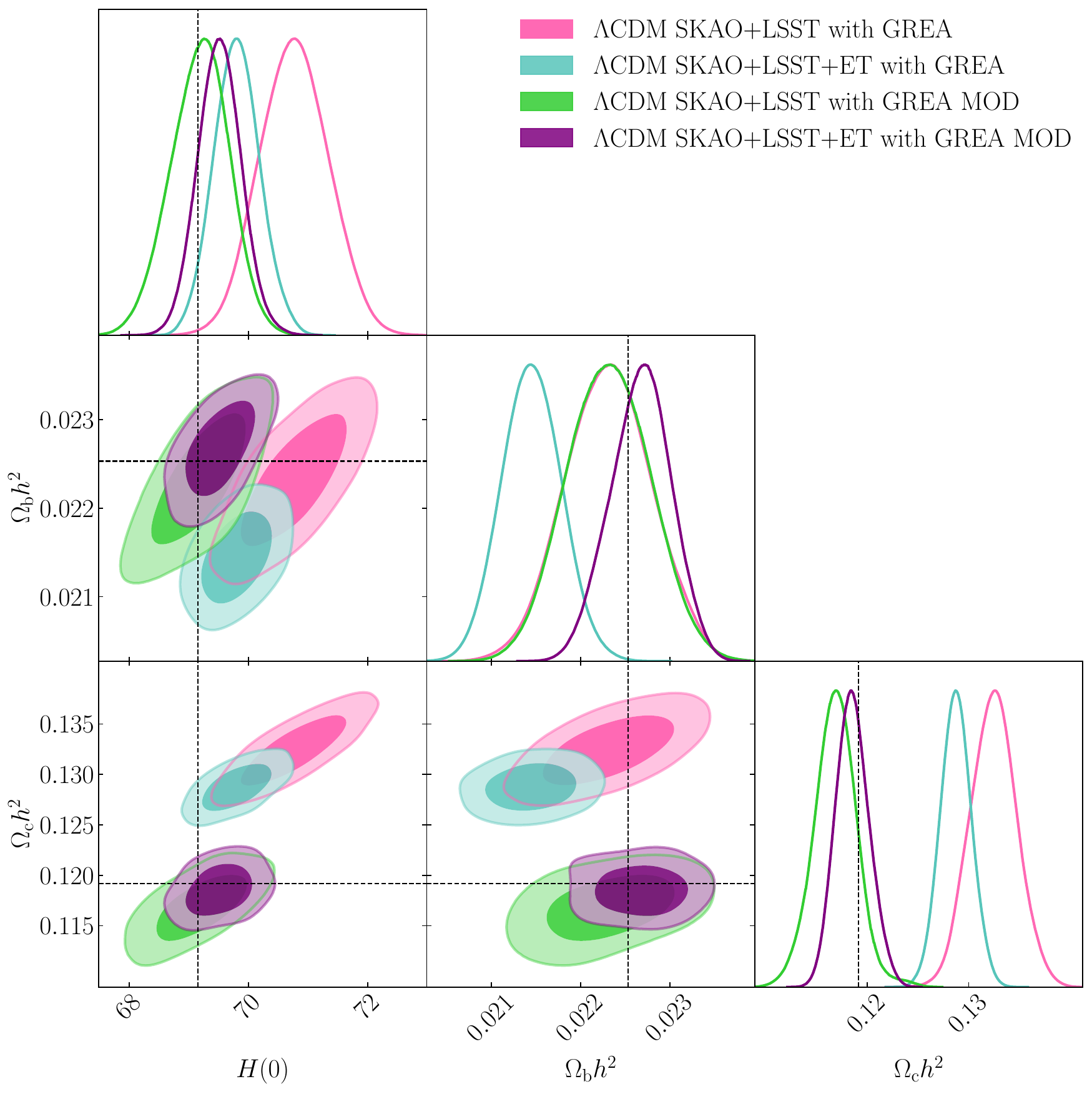}
    	\caption{Posterior distribution of GREA and modified GREA parameter from $\Lambda$CDM SKAO+LSST (pink and green), $\Lambda$CDM SKAO+LSST+ET (cyan and purple).
        The dashed lines correspond to the parameters value that we used to generate the mock datasets.}
        \label{fig:figura_6}
    \end{figure}
\end{center}

\autoref{fig:figura_6} (pink and cyan contours) shows the results of the analysis of these datasets with the GREA theory.
We find that the recovered values for the primary cosmological parameter, specifically $H(0)$, $\Omega_bh^2$ and $\Omega_{c}h^2$, are systematically shifted with respect to their $\Lambda$CDM fiducial values.
This behavior indicates that, in the absence of an explicit cosmological constant, the GREA model compensates by adjusting the matter density and the expansion rate in order to reproduce the late time acceleration present in the $\Lambda$CDM mock data.

The inclusion of GW further enhances this discrepancy.
Although GW observations provide precise measurements of the luminosity distance, they tighten the constraints around configurations that deviate from the fiducial $\Lambda$CDM model.
Consequently, the posterior distributions exhibit more pronounced shifts, highlighting the difficulty of reproducing a $\Lambda$CDM expansion history within the GREA framework without inducing correlated shifts in the inferred physical parameters.

In contrast, the modified GREA framework (\autoref{fig:figura_6}, green and purple contours) substantially alleviates these systematic shifts.
By introducing an explicit dark energy component, the model no longer requires large compensatory shifts in the standard cosmological parameters in order to reproduce $\Lambda$CDM mock data.

We find that the inferred values of $H(0)$, $\Omega_bh^2$, and $\Omega_{c} h^2$ recover the fiducial $\Lambda$CDM values within $1\sigma$. 
In this scenario, the additional dark energy parameter effectively accounts for the late-time acceleration, allowing the remaining physical components to remain consistent with the underlying cosmology.

This behavior indicates that the modified GREA framework can consistently accommodate a $\Lambda$CDM-like expansion history when supported by the data, while the pure GREA framework exhibits significant parameter shifts due to the absence of an explicit dark energy term.

\subsubsection{Bayesian model comparison}
\label{subsubsec:LCDMmock_bayesiancomparison}

The Bayesian evidence for the $\Lambda$CDM, GREA and modified GREA is reported in \autoref{table:evidence_mockdataLCDM}.

For the $\Lambda$CDM SKAO+LSST mock data combination, $\Lambda$CDM is strongly favored over both alternative frameworks.
The GREA model is strongly disfavoured, with $\Delta \text{log} Z=-10.16$, while the modified GREA shows a significant improvement, reducing the evidence difference to $\Delta \text{log} Z= -6.21$.
However, both alternative models are disfavoured with respect to $\Lambda$CDM.

With the inclusion of GW mock data, $\Lambda$CDM SKAO+LSST+ET, the evidence difference increases to $\Delta \text{log} Z=-12.39$ for GREA and remains $\Delta \text{log} Z= -4.95$ for modified GREA.
This indicates that the inclusion of future GW observations can significantly enhance the statistical power to identify the correct cosmological model and penalize alternative descriptions that introduce unnecessary complexity.

Overall, these results demonstrate that when the true cosmology is $\Lambda$CDM, both GREA and its modified extension are statistically disfavoured in a Bayesian model comparison.
Although the modified GREA performs better than the original formulation, it remains penalized by the presence of an additional free parameter that is not required by the data.

\begingroup 
    \setlength{\tabcolsep}{10pt}
    \renewcommand{\arraystretch}{1.5} 
    \begin{table}
        \centering
        \begin{tabular}{ c c c c }  
            Model & Data combination & $\Delta \text{log} Z$\\
            \hline \hline  
            $\Lambda$CDM & $\Lambda$CDM SKAO+LSST & 0 \\
            GREA & $\Lambda$CDM SKAO+LSST & -10.163\\
            Modified GREA & $\Lambda$CDM SKAO+LSST & \cdl{-6.21} \\ 
            $\Lambda$CDM & $\Lambda$CDM SKAO+LSST+ET & 0\\
            GREA & $\Lambda$CDM SKAO+LSST+ET & -12.399\\
            Modified GREA & $\Lambda$CDM SKAO+LSST+ET & \cdl{-4.95} \\
            \hline \hline  
        \end{tabular}  
        \caption{Bayesian evidence and Bayes factor for $\Lambda$CDM, GREA and modified GREA obtained from the analysis of $\Lambda$CDM SKAO+LSST and $\Lambda$CDM SKAO+LSST+ET. The Bayes factor is computed respect to $\Lambda$CDM as $\Delta \text{log} Z= \text{log}Z_{model}-\text{log}Z_{\Lambda CDM}$.}
        \label{table:evidence_mockdataLCDM}
    \end{table}
\endgroup

\newpage
\section{Conclusions}
\label{sec:conclusion}
In this work, we investigated the General Relativistic Entropic Acceleration (GREA) framework as an alternative to the cosmological constant ($\Lambda$) for explaining late-time cosmic acceleration.
In this scenario, the accelerated expansion arises from entropy production associated with the cosmological horizon.
This approach is motivated by the theoretical challenges of $\Lambda$CDM, including the cosmological constant problem, and by the current observational tensions, most notably the discrepancy in the present-day expansion rate inferred from early- and late-time probes.

Our analysis first constrained the GREA model using a comprehensive suite of background observations, including baryon acoustic oscillations, type Ia supernovae, compressed CMB information, and cosmic chronometers.
We find that when CMB information is included, $\Lambda$CDM is statistically preferred in a Bayesian model comparison, while GREA is highly competitive in low-redshift datasets.
However, the GREA framework accommodates expansion histories that shift the inferred value of $H(0)$ toward those measured by local probes.
In this regime, the geometric horizon parameter $\sqrt{-k}\eta_0$ plays a central role in controlling the late-time dynamics.
We also considered a phenomenological extension of the framework including an additional dark energy component, $\Omega_{de}$, which allows the recovery of a $\Lambda$CDM-like expansion history as a limiting case and provides a controlled interpolation between entropic- and dark-energy-driven acceleration. 
The analysis of this extended scenario shows that the extra degree of freedom introduces significant degeneracies with the main GREA parameter, $\sqrt{-k}\eta_0$.
The posterior distribution of $\Omega_{de}$ is highly asymmetric and sensitive to the adopted prior on $\sqrt{-k}\eta_0$. 
For this reason, in all the analysis of modified GREA we adopt an extended prior range for $\sqrt{-k}\eta_0$ compared to that used in GREA.

To evaluate the sensitivity of future observations, we performed a forecast analysis using mock datasets representative of next-generation surveys such as SKAO, LSST, and ET.
When the fiducial cosmology is described by GREA, future multi-probe datasets are expected to recover the underlying cosmology.
In particular, the inclusion of gravitational-wave standard sirens not only tightens parameter constraints but also significantly enhances the Bayesian discrimination between GREA and $\Lambda$CDM, disfavoring the introduction in modified GREA of an unnecessary dark energy component when the true cosmology is entropic in origin.

Conversely, when the fiducial cosmology is $\Lambda$CDM, the GREA theory compensates for the absence of an explicit cosmological constant through correlated shifts in the inferred matter density and expansion rate, leading to statistically disfavored fits once high-precision distance measurements are included.
The modified GREA framework is able to reproduce the $\Lambda$CDM expansion history through a non-zero $\Omega_{de}$, but this additional degree of freedom is penalized in Bayesian evidence when not required by the data.

Overall, our results indicate that future high-precision observations, and in particular gravitational wave standard sirens, will provide a powerful and independent test of entropic cosmological scenarios.
The combination of parameter inference and Bayesian model comparison will be essential to determine whether horizon thermodynamics can offer a viable alternative to a fundamental cosmological constant or if the observed tensions necessitate alternative modifications to the standard cosmological model.

\mmt{Future extensions of this work should incorporate linear perturbation and full CMB likelihood analyses, allowing for a more comprehensive assessment of structure formation and growth within the GREA framework. Such an analysis would allow to extract more information from CMB data, and the study of perturbation could disentangle GREA from models that yields similar expansion history. In addition to this, a full CMB analysis would also allow to remove any possible bias introduced by the use of the compressed CMB likelihood, whose information is obtained within a $\Lambda$CDM assumption. Removing such a possible bias will allow to clarify if the significant preference of CMB data for $\Lambda$CDM is indeed a feature of the data or is due to assumptions hidden within the CMB compressed likelihood. A step in this direction has been taken by \citet{garciabellido2024darkenergypredictionsgrea}, with a first derivation of perturbation evolution in GREA. However, such predictions are still not implemented in a Boltzmann solver, which is needed to exploit the full CMB analysis pipeline.}

\begingroup 
    \setlength{\tabcolsep}{5pt}
    \renewcommand{\arraystretch}{1.5} 
    \begin{table}[H]
        \hspace*{-0.8cm}
        \begin{tabular}{ | c | c c c c c c c| }  
        \hline
            Dataset & Model & $H(0)$ & $H_0$ &$\Omega_m$ & $\sqrt{-k}\eta_0$ & $\Omega_{de}$ & $\Omega_{GREA}$\\
            \hline \hline  

             & $\Lambda$CDM & $69.32\pm 0.25$& $69.32\pm 0.25$  & $0.2941\pm 0.0032$ & - & $0.7058\pm 0.0032$ & - \\
            DESI+P-ACT+Pantheon+ & GREA & $70.56\pm 0.46$ & $70.51\pm 0.57$ & $0.2861\pm 0.0035$ & $3.640\pm 0.019$ & - & $0.7148\pm 0.0037$ \\
             & GREA Mod & $69.74\pm 0.42$ & $69.91\pm 0.53$ & $0.2920^{+0.0035}_{-0.0031}$ & $7.4\pm 1.6$ & $0.695^{+0.014}_{\cdl{-}0.0018}$ & $0.0097^{\cdl{+}0.0027}_{-0.010}$ \\
            \hline
             & $\Lambda$CDM & $69.33\pm 0.25$ & $69.33\pm 0.25$ & $0.2939\pm 0.0032$ & - & $0.7060\pm 0.0032$ & -\\
            DESI+P-ACT+CC+Pantheon+ & GREA & $70.49\pm 0.45$ & $70.41\pm 0.55$ & $0.2862\pm 0.0035$ & $3.638\pm 0.019$ & - & $0.7154\pm 0.0037$\\
             & GREA Mod & $69.61\pm 0.38$  & $69.72\pm 0.49$& $0.2924\pm 0.0034$ & $7.0\pm 1.6$ & $0.688^{+0.024}_{\cdl{-}0.0066}$ & $0.018^{\cdl{+}0.010}_{-0.018}$ \\
            \hline
             & $\Lambda$CDM & $69.73\pm 0.44$ & $69.73\pm 0.44$& $0.2994\pm 0.0068$ & - & $0.7005\pm 0.0068$ & - \\
            DESI+CC+Pantheon+ & GREA & $69.55\pm 0.92$ & $68.5\pm 1.5$ & $0.3103\pm 0.0084$ & $3.487\pm 0.060$ & - & $0.711\pm 0.010$\\
             & GREA Mod & $69.29\pm 0.85$ & $68.8\pm 1.5$& $0.3027^{+0.0094}_{-0.011}$ & $6.7\pm 1.9$ & $0.654^{+0.070}_{+0.027}$ & $0.054^{\cdl{+}0.034}_{-0.055}$\\
            \hline
             & $\Lambda$CDM & $70.32\pm 0.50$ & $70.32\pm 0.50$ & $0.3107\pm 0.0079$ & - & $0.6892\pm 0.0079$ & - \\
            \raisebox{0.75\normalbaselineskip}{DESI+Pantheon+} & GREA & $71.1\pm 1.1$ & $70.8\pm 1.8$& $0.3160\pm 0.0088$ & $3.514\pm 0.063$ & - & $0.690\pm 0.012$ \\
            \hline
            & $\Lambda$CDM & $69.56\pm 0.41$ & $69.56\pm 0.41$ & $0.2910\pm 0.0053$ & - & $0.7089\pm 0.0053$ & - \\
            P-ACT+CC+Pantheon+ & GREA & $70.49\pm 0.44$ & $70.40\pm 0.54$ & $0.2863\pm 0.0034$ & $3.638\pm 0.018$ & - & $0.7154\pm 0.0037$ \\
            & GREA Mod & $69.51\pm 0.36$ & $71.8\pm1.3$ & $0.2923\pm 0.0036$ & $7.2\pm 1.6$ & $0.692^{+0.018}_{\cdl{-}0.0048}$ & $0.0130^{\cdl{+}0.0097}_{-0.013}$ \\
            \hline
            & $\Lambda$CDM & $68.17\pm 0.42$ & $68.17\pm 0.42$ & $0.2710\pm 0.0026$ & - & $0.7289\pm 0.0026$ & - \\
            \raisebox{0.75\normalbaselineskip}{GREA SKAO+LSST} & GREA & $69.91\pm 0.55$ & $69.91\pm 0.69$ & $0.2900\pm 0.0026$ & $3.608\pm 0.020$ & - & $0.7162\pm 0.0038$ \\
            \hline
             & $\Lambda$CDM & $69.92\pm 0.23$ & $69.92\pm 0.23$ & $0.2740\pm 0.0025$ & - & $0.7259\pm 0.0025$ & - \\
            GREA SKAO+LSST+ET & GREA & $70.28\pm 0.37$ & $70.00\pm 0.53$ & $0.2902\pm 0.0026$ & $3.609\pm 0.020$ & - & $0.7155\pm 0.0037$ \\
             & GREA Mod &  $70.23\pm 0.33$ &  $70.01\pm0.48$ &$0.2865^{+0.0036}_{-0.0028}$ & $3.743^{+0.043}_{-0.10}$ & $0.146^{+0.051}_{-0.098}$ & $0.571^{+0.096}_{-0.051}$ \\
            \hline
             & $\Lambda$CDM & $69.00\pm 0.43$& $69.00\pm 0.43$ & $0.2928\pm 0.0028$ & - & $0.7071\pm 0.0028$ & - \\
            $\Lambda$CDM SKAO+LSST & GREA & $70.77\pm 0.57$ & $70.63\pm 0.72$  & $0.3090\pm 0.0029$ & $3.549\pm 0.020$ & - & $0.6937\pm 0.0037$ \\
             & GREA Mod &  $69.19^{+0.52}_{-0.46}$  &  $69.35\pm0.62$& $0.2908^{+0.0030}_{-0.0034}$ & $7.2\pm 1.7$ & $0.690^{+0.022}_{\cdl{-}0.0072}$ & $0.016^{\cdl{+}0.012}_{-0.016}$\\
            \hline
             & $\Lambda$CDM & $69.36\pm 0.25$ & $69.36\pm 0.25$& $0.2933\pm 0.0027 $ & - & $0.7067\pm 0.0027$ & - \\
            $\Lambda$CDM SKAO+LSST+ET & GREA & $69.80\pm 0.38$ & $69.61\pm 0.54$ & $0.3083\pm 0.0028 $ & $3.547\pm 0.019$ & - & $0.6955\pm 0.0036$ \\
             & GREA Mod & $69.51\pm 0.36 $ & $69.65\pm 0.31 $ & $0.2923\pm 0.0036$ & $7.2\pm 1.6$ & $0.692^{+0.018}_{\cdl{-}0.0048}$ & $0.0130^{\cdl{+}0.0097}_{-0.013}$  \\
            \hline \hline  
        \end{tabular}  
        \caption{Best fit values and $68\%$ confidence levels for the cosmological parameters obtained from different data combinations and models. \cdl{We highlight that by construction in the $\Lambda$CDM model $H(0)=H_0$, while these two do not coincide in the GREA and GREA Modified scenarios.}}
        \label{table:results}
    \end{table}
\endgroup

\section*{Acknowledgments}
I.G. acknowledges financial support from the ICSC - Centro Nazionale di Ricerca in HPC, Big Data e Quantum Computing, CN00000013, spoke 10 “quantum computing”, CUP C53C22000350006.
M.M. acknowledges funding by the Agenzia Spaziale Italiana (\textsc{asi}) under agreement n. 2024-10-HH.0 and support from INFN/Euclid Sezione di Roma. C.D.L. and M.M. acknowledge financial support from Sapienza Università di Roma, provided through Progetti Medi 2021 (Grant No. RM12117A51D5269B).
\noindent This work made use of Melodie, a computing infrastructure funded by the same project.
\noindent The authors thank R. Calderon for useful discussions.

\newpage
\bibliographystyle{abbrvnat}
\bibliography{oja_template}

@article{Garc_a_Bellido_2021,
    author = "Garcia-Bellido, Juan and Espinosa-Portales, Llorenc",
    title = "{Cosmic acceleration from first principles}",
    eprint = "2106.16014",
    archivePrefix = "arXiv",
    primaryClass = "gr-qc",
    reportNumber = "IFT-UAM/CSIC-21-75",
    doi = "10.1016/j.dark.2021.100892",
    journal = "Phys. Dark Univ.",
    volume = "34",
    pages = "100892",
    year = "2021"
}

@article{Arjona_2022,
    author = "Arjona, Rub{\'e}n and Espinosa-Portales, Llorenc and Garc{\'\i}a-Bellido, Juan and Nesseris, Savvas",
    title = "{A GREAT model comparison against the cosmological constant}",
    eprint = "2111.13083",
    archivePrefix = "arXiv",
    primaryClass = "astro-ph.CO",
    reportNumber = "IFT-UAM/CSIC-2021-136",
    doi = "10.1016/j.dark.2022.101029",
    journal = "Phys. Dark Univ.",
    volume = "36",
    pages = "101029",
    year = "2022"
}

@article{Scolnic_2022,
    author = "Scolnic, Dan and others",
    title = "{The Pantheon+ Analysis: The Full Data Set and Light-curve Release}",
    eprint = "2112.03863",
    archivePrefix = "arXiv",
    primaryClass = "astro-ph.CO",
    doi = "10.3847/1538-4357/ac8b7a",
    journal = "Astrophys. J.",
    volume = "938",
    number = "2",
    pages = "113",
    year = "2022"
}

@article{Moresco_2022,
    author = "Moresco, Michele and others",
    title = "{Unveiling the Universe with emerging cosmological probes}",
    eprint = "2201.07241",
    archivePrefix = "arXiv",
    primaryClass = "astro-ph.CO",
    doi = "10.1007/s41114-022-00040-z",
    journal = "Living Rev. Rel.",
    volume = "25",
    number = "1",
    pages = "6",
    year = "2022"
}

@article{Moresco_2016,
    author = "Moresco, Michele and Pozzetti, Lucia and Cimatti, Andrea and Jimenez, Raul and Maraston, Claudia and Verde, Licia and Thomas, Daniel and Citro, Annalisa and Tojeiro, Rita and Wilkinson, David",
    title = "{A 6{\%} measurement of the Hubble parameter at $z\sim0.45$: direct evidence of the epoch of cosmic re-acceleration}",
    eprint = "1601.01701",
    archivePrefix = "arXiv",
    primaryClass = "astro-ph.CO",
    doi = "10.1088/1475-7516/2016/05/014",
    journal = "JCAP",
    volume = "05",
    pages = "014",
    year = "2016"
}

@article{Torrado_2021,
    author = "Torrado, Jesus and Lewis, Antony",
    title = "{Cobaya: Code for Bayesian Analysis of hierarchical physical models}",
    eprint = "2005.05290",
    archivePrefix = "arXiv",
    primaryClass = "astro-ph.IM",
    reportNumber = "TTK-20-15",
    doi = "10.1088/1475-7516/2021/05/057",
    journal = "JCAP",
    volume = "05",
    pages = "057",
    year = "2021"
}

@article{nautilus,
    author = "Lange, Johannes U.",
    title = "{nautilus: boosting Bayesian importance nested sampling with deep learning}",
    eprint = "2306.16923",
    archivePrefix = "arXiv",
    primaryClass = "astro-ph.IM",
    doi = "10.1093/mnras/stad2441",
    journal = "Mon. Not. Roy. Astron. Soc.",
    volume = "525",
    number = "2",
    pages = "3181--3194",
    year = "2023"
}

@article{Ryan_2020,
   title={Gamma-Ray Burst Afterglows in the Multimessenger Era: Numerical Models and Closure Relations},
   volume={896},
   ISSN={1538-4357},
   url={http://dx.doi.org/10.3847/1538-4357/ab93cf},
   DOI={10.3847/1538-4357/ab93cf},
   number={2},
   journal={The Astrophysical Journal},
   publisher={American Astronomical Society},
   author={Ryan, Geoffrey and Eerten, Hendrik van and Piro, Luigi and Troja, Eleonora},
   year={2020},
   month=jun, pages={166} }

@article{Adame_2025,
    author = "Adame, A. G. and others",
    collaboration = "DESI",
    title = "{DESI 2024 VI: cosmological constraints from the measurements of baryon acoustic oscillations}",
    eprint = "2404.03002",
    archivePrefix = "arXiv",
    primaryClass = "astro-ph.CO",
    reportNumber = "FERMILAB-PUB-24-0154-PPD",
    doi = "10.1088/1475-7516/2025/02/021",
    journal = "JCAP",
    volume = "02",
    pages = "021",
    year = "2025"
}

@article{Efstathiou_2021,
    author = "Efstathiou, George",
    title = "{To H0 or not to H0?}",
    eprint = "2103.08723",
    archivePrefix = "arXiv",
    primaryClass = "astro-ph.CO",
    doi = "10.1093/mnras/stab1588",
    journal = "Mon. Not. Roy. Astron. Soc.",
    volume = "505",
    number = "3",
    pages = "3866--3872",
    year = "2021"
}

@article{heavens2010statisticaltechniquescosmology,
    author = "Heavens, Alan",
    title = "{Statistical techniques in cosmology}",
    eprint = "0906.0664",
    doi = {10.48550/arXiv.0906.0664},
    archivePrefix = "arXiv",
    primaryClass = "astro-ph.CO",
    month = "6",
    year = "2009"
}

@article{Nesseris_2013,
    author = "Nesseris, Savvas and Garcia-Bellido, Juan",
    title = "{Is the Jeffreys' scale a reliable tool for Bayesian model comparison in cosmology?}",
    eprint = "1210.7652",
    archivePrefix = "arXiv",
    primaryClass = "astro-ph.CO",
    reportNumber = "IFT-UAM-CSIC-12-95",
    doi = "10.1088/1475-7516/2013/08/036",
    journal = "JCAP",
    volume = "08",
    pages = "036",
    year = "2013"
}

@article{Aizpuru_2021,
    author = "Aizpuru, Andoni and Arjona, Rub{\'e}n and Nesseris, Savvas",
    title = "{Machine learning improved fits of the sound horizon at the baryon drag epoch}",
    eprint = "2106.00428",
    archivePrefix = "arXiv",
    primaryClass = "astro-ph.CO",
    reportNumber = "IFT-UAM/CSIC-21-67",
    doi = "10.1103/PhysRevD.104.043521",
    journal = "Phys. Rev. D",
    volume = "104",
    number = "4",
    pages = "043521",
    year = "2021"
}

@article{lodha2025extendeddarkenergyanalysis,
    author = "Lodha, K. and others",
    collaboration = "DESI",
    title = "{Extended dark energy analysis using DESI DR2 BAO measurements}",
    eprint = "2503.14743",
    archivePrefix = "arXiv",
    primaryClass = "astro-ph.CO",
    reportNumber = "FERMILAB-PUB-25-0164-PPD",
    doi = "10.1103/w4c6-1r5j",
    journal = "Phys. Rev. D",
    volume = "112",
    number = "8",
    pages = "083511",
    year = "2025"
}

@article{Abdul_Karim_2025,
    author = "Abdul Karim, M. and others",
    collaboration = "DESI",
    title = "{DESI DR2 results. II. Measurements of baryon acoustic oscillations and cosmological constraints}",
    eprint = "2503.14738",
    archivePrefix = "arXiv",
    primaryClass = "astro-ph.CO",
    reportNumber = "FERMILAB-PUB-25-0169-PPD",
    doi = "10.1103/tr6y-kpc6",
    journal = "Phys. Rev. D",
    volume = "112",
    number = "8",
    pages = "083515",
    year = "2025"
}

@article{Lemos_2023,
    author = "Lemos, Pablo and Lewis, Antony",
    title = "{CMB constraints on the early Universe independent of late-time cosmology}",
    eprint = "2302.12911",
    archivePrefix = "arXiv",
    primaryClass = "astro-ph.CO",
    doi = "10.1103/PhysRevD.107.103505",
    journal = "Phys. Rev. D",
    volume = "107",
    number = "10",
    pages = "103505",
    year = "2023"
}

@article{calderon2025constraininggreaalternativetheory,
    author = "Calderon, R. and others",
    title = "{Constraining GREA, an alternative theory accounting for the present cosmic acceleration}",
    eprint = "2509.21491",
    doi = {10.48550/arXiv.2509.21491},
    archivePrefix = "arXiv",
    primaryClass = "astro-ph.CO",
    reportNumber = "DESI-2024-0464, IFT-UAM/CSIC-25-99, FERMILAB-PUB-25-0737-PPD",
    month = "9",
    year = "2025"
}

@article{louis2025atacamacosmologytelescopedr6,
    author = "Louis, Thibaut and others",
    collaboration = "Atacama Cosmology Telescope",
    title = "{The Atacama Cosmology Telescope: DR6 power spectra, likelihoods and {\ensuremath{\Lambda}}CDM parameters}",
    eprint = "2503.14452",
    archivePrefix = "arXiv",
    primaryClass = "astro-ph.CO",
    reportNumber = "FERMILAB-PUB-25-0071-PPD",
    doi = "10.1088/1475-7516/2025/11/062",
    journal = "JCAP",
    volume = "11",
    pages = "062",
    year = "2025"
}

@article{garciabellido2024darkenergypredictionsgrea,
    author = "Garcia-Bellido, Juan",
    title = "{Dark Energy predictions from GREA: Background and linear perturbation theory}",
    eprint = "2405.02895",
    archivePrefix = "arXiv",
    primaryClass = "astro-ph.CO",
    reportNumber = "IFT-UAM/CSIC-24-69",
    doi = "10.1016/j.dark.2024.101533",
    journal = "Phys. Dark Univ.",
    volume = "45",
    pages = "101533",
    year = "2024"
}

@article{deleo2025distinguishingdistancedualitybreaking,
    author = "De Leo, Chiara and Martinelli, Matteo and D'Agostino, Rocco and Gianfagna, Giulia and Martins, Carlos J. A. P.",
    title = "{Distinguishing distance duality breaking models using electromagnetic and gravitational waves measurements}",
    eprint = "2505.13613",
    archivePrefix = "arXiv",
    primaryClass = "astro-ph.CO",
    reportNumber = "ET-0147A-25",
    doi = "10.1088/1475-7516/2025/11/001",
    journal = "JCAP",
    volume = "11",
    pages = "001",
    year = "2025"
}

@article{fazzari2025investigatingfrinflationbackgroundevolution,
    author = "Fazzari, Elisa and De Leo, Chiara and Montani, Giovanni and Martinelli, Matteo and Melchiorri, Alessandro and Ca{\~n}as-Herrera, Guadalupe",
    title = "{Investigating $f(R)$-Inflation: background evolution and constraints}",
    eprint = "2507.13890",
    doi = {10.48550/arXiv.2507.13890},
    archivePrefix = "arXiv",
    primaryClass = "astro-ph.CO",
    month = "7",
    year = "2025"
}

@article{2020,
    author = "Bacon, David J. and others",
    collaboration = "SKA",
    title = "{Cosmology with Phase 1 of the Square Kilometre Array: Red Book 2018: Technical specifications and performance forecasts}",
    eprint = "1811.02743",
    archivePrefix = "arXiv",
    primaryClass = "astro-ph.CO",
    doi = "10.1017/pasa.2019.51",
    journal = "Publ. Astron. Soc. Austral.",
    volume = "37",
    pages = "e007",
    year = "2020"
}

@article{Bianco_2021,
    author = "Bianco, Federica B. and others",
    title = "{Optimization of the Observing Cadence for the Rubin Observatory Legacy Survey of Space and Time: A Pioneering Process of Community-focused Experimental Design}",
    eprint = "2108.01683",
    archivePrefix = "arXiv",
    primaryClass = "astro-ph.IM",
    reportNumber = "FERMILAB-PUB-22-003-SCD",
    doi = "10.3847/1538-4365/ac3e72",
    journal = "Astrophys. J. Supp.",
    volume = "258",
    number = "1",
    pages = "1",
    year = "2022"
}

@article{Gong_2010,
    author = "Gong, Yan and Cooray, Asantha and Chen, Xuelei",
    title = "{Cosmology with Photometric Surveys of Type Ia Supernovae}",
    eprint = "0909.2692",
    archivePrefix = "arXiv",
    primaryClass = "astro-ph.CO",
    doi = "10.1088/0004-637X/709/2/1420",
    journal = "Astrophys. J.",
    volume = "709",
    pages = "1420--1428",
    year = "2010"
}

@article{Astier_2014,
    author = "Astier, P. and others",
    title = "{Extending the supernova Hubble diagram to z $\sim$ 1.5 with the Euclid space mission}",
    eprint = "1409.8562",
    archivePrefix = "arXiv",
    primaryClass = "astro-ph.CO",
    doi = "10.1051/0004-6361/201423551",
    journal = "Astron. Astrophys.",
    volume = "572",
    pages = "A80",
    year = "2014"
}

@article{Holz_2005,
    author = "Holz, Daniel E. and Hughes, Scott A.",
    title = "{Using gravitational-wave standard sirens}",
    eprint = "astro-ph/0504616",
    archivePrefix = "arXiv",
    doi = "10.1086/431341",
    journal = "Astrophys. J.",
    volume = "629",
    pages = "15--22",
    year = "2005"
}

@article{Nissanke_2010,
    author = "Nissanke, Samaya and Holz, Daniel E. and Hughes, Scott A. and Dalal, Neal and Sievers, Jonathan L.",
    title = "{Exploring short gamma-ray bursts as gravitational-wave standard sirens}",
    eprint = "0904.1017",
    archivePrefix = "arXiv",
    primaryClass = "astro-ph.CO",
    doi = "10.1088/0004-637X/725/1/496",
    journal = "Astrophys. J.",
    volume = "725",
    pages = "496--514",
    year = "2010"
}

@article{Martinelli_2022,
    author = "Martinelli, Matteo and Scarcella, Francesca and Hogg, Natalie B. and Kavanagh, Bradley J. and Gaggero, Daniele and Fleury, Pierre",
    title = "{Dancing in the dark: detecting a population of distant primordial black holes}",
    eprint = "2205.02639",
    archivePrefix = "arXiv",
    primaryClass = "astro-ph.CO",
    reportNumber = "IFT-UAM/CSIC-22-50",
    doi = "10.1088/1475-7516/2022/08/006",
    journal = "JCAP",
    volume = "08",
    number = "08",
    pages = "006",
    year = "2022"
}

@article{Dupletsa_2023,
    author = "Dupletsa, Ulyana and Harms, Jan and Banerjee, Biswajit and Branchesi, Marica and Goncharov, Boris and Maselli, Andrea and Oliveira, Ana Carolina Silva and Ronchini, Samuele and Tissino, Jacopo",
    title = "{gwfish: A simulation software to evaluate parameter-estimation capabilities of gravitational-wave detector networks}",
    eprint = "2205.02499",
    archivePrefix = "arXiv",
    primaryClass = "gr-qc",
    doi = "10.1016/j.ascom.2022.100671",
    journal = "Astron. Comput.",
    volume = "42",
    pages = "100671",
    year = "2023"
}

@article{abac2025scienceeinsteintelescope,
    author = "Abac, Adrian and others",
    collaboration = "ET",
    title = "{The Science of the Einstein Telescope}",
    doi = {10.48550/arXiv.2503.12263},
    eprint = "2503.12263",
    archivePrefix = "arXiv",
    primaryClass = "gr-qc",
    reportNumber = "ET-0036C-25",
    month = "3",
    year = "2025"
}

@article{Mathews_2017,
     author = "Mathews, Grant J. and Kusakabe, Motohiko and Kajino, Toshitaka",
    title = "{Introduction to Big Bang Nucleosynthesis and Modern Cosmology}",
    eprint = "1706.03138",
    archivePrefix = "arXiv",
    primaryClass = "astro-ph.CO",
    doi = "10.1142/S0218301317410014",
    journal = "Int. J. Mod. Phys. E",
    volume = "26",
    number = "08",
    pages = "1741001",
    year = "2017"
}

@article{Espinosa_Portal_s_2021,
    author = "Espinosa-Portales, Llorenc and Garcia-Bellido, Juan",
    title = "{Covariant formulation of non-equilibrium thermodynamics in General Relativity}",
    eprint = "2106.16012",
    archivePrefix = "arXiv",
    primaryClass = "gr-qc",
    reportNumber = "IFT-UAM/CSIC-21-74",
    doi = "10.1016/j.dark.2021.100893",
    journal = "Phys. Dark Univ.",
    volume = "34",
    pages = "100893",
    year = "2021"
}

@article{PhysRevD.15.2752,
    author = "Gibbons, G. W. and Hawking, S. W.",
    title = "{Action Integrals and Partition Functions in Quantum Gravity}",
    reportNumber = "PRINT-76-0995 (CAMBRIDGE)",
    doi = "10.1103/PhysRevD.15.2752",
    journal = "Phys. Rev. D",
    volume = "15",
    pages = "2752--2756",
    year = "1977"
}

@article{Riess_2022,
    author = "Riess, Adam G. and others",
    title = "{A Comprehensive Measurement of the Local Value of the Hubble Constant with 1 km s$^{-1}$ Mpc$^{-1}$ Uncertainty from the Hubble Space Telescope and the SH0ES Team}",
    eprint = "2112.04510",
    archivePrefix = "arXiv",
    primaryClass = "astro-ph.CO",
    doi = "10.3847/2041-8213/ac5c5b",
    journal = "Astrophys. J. Lett.",
    volume = "934",
    number = "1",
    pages = "L7",
    year = "2022"
}

@article{Scolnic_2018,
    author = "Scolnic, D. M. and others",
    collaboration = "Pan-STARRS1",
    title = "{The Complete Light-curve Sample of Spectroscopically Confirmed SNe Ia from Pan-STARRS1 and Cosmological Constraints from the Combined Pantheon Sample}",
    eprint = "1710.00845",
    archivePrefix = "arXiv",
    primaryClass = "astro-ph.CO",
    doi = "10.3847/1538-4357/aab9bb",
    journal = "Astrophys. J.",
    volume = "859",
    number = "2",
    pages = "101",
    year = "2018"
}

@article{Robert_2009,
   title={Harold Jeffreys’s Theory of Probability Revisited},
   volume={24},
   ISSN={0883-4237},
   doi={10.1214/09-sts284},
   number={2},
   journal={Statistical Science},
   publisher={Institute of Mathematical Statistics},
   author={Robert, Christian P. and Chopin, Nicolas and Rousseau, Judith},
   year={2009},
   month=may }

@book{jeffreys1998theory,
    author = "Jeffreys, Harold",
    title = "{The Theory of Probability}",
    isbn = "978-0-19-850368-2, 978-0-19-853193-7",
    series = "Oxford Classic Texts in the Physical Sciences",
    year = "1939"
}

@article{Bahamonde_2018,
    author = {Bahamonde, Sebastian and B{\"o}hmer, Christian G. and Carloni, Sante and Copeland, Edmund J. and Fang, Wei and Tamanini, Nicola},
    title = "{Dynamical systems applied to cosmology: dark energy and modified gravity}",
    eprint = "1712.03107",
    archivePrefix = "arXiv",
    primaryClass = "gr-qc",
    doi = "10.1016/j.physrep.2018.09.001",
    journal = "Phys. Rept.",
    volume = "775-777",
    pages = "1--122",
    year = "2018"
}

@article{Di_Valentino_2021,
    author = "Di Valentino, Eleonora and Mena, Olga and Pan, Supriya and Visinelli, Luca and Yang, Weiqiang and Melchiorri, Alessandro and Mota, David F. and Riess, Adam G. and Silk, Joseph",
    title = "{In the realm of the Hubble tension{\textemdash}a review of solutions}",
    eprint = "2103.01183",
    archivePrefix = "arXiv",
    primaryClass = "astro-ph.CO",
    reportNumber = "IPPP/20/108",
    doi = "10.1088/1361-6382/ac086d",
    journal = "Class. Quant. Grav.",
    volume = "38",
    number = "15",
    pages = "153001",
    year = "2021"
}

@article{garciabellido2025greadarkenergyholographic,
    author = "Garc{\'\i}a-Bellido, Juan",
    title = "{GREA and Dark Energy: A holographic dual}",
    eprint = "2511.19546",
    doi = {10.48550/arXiv.2511.19546},
    archivePrefix = "arXiv",
    primaryClass = "gr-qc",
    reportNumber = "IFT-UAM/CSIC-25-149",
    month = "11",
    year = "2025"
}

@article{Abdalla_2022,
    author = "Abdalla, Elcio and others",
    title = "{Cosmology intertwined: A review of the particle physics, astrophysics, and cosmology associated with the cosmological tensions and anomalies}",
    eprint = "2203.06142",
    archivePrefix = "arXiv",
    primaryClass = "astro-ph.CO",
    reportNumber = "FERMILAB-CONF-22-192-SCD",
    doi = "10.1016/j.jheap.2022.04.002",
    journal = "JHEAp",
    volume = "34",
    pages = "49--211",
    year = "2022"
}

@article{Velten_2014,
    author = "Velten, H. E. S. and vom Marttens, R. F. and Zimdahl, W.",
    title = "{Aspects of the cosmological {\textquotedblleft}coincidence problem{\textquotedblright}}",
    eprint = "1410.2509",
    archivePrefix = "arXiv",
    primaryClass = "astro-ph.CO",
    doi = "10.1140/epjc/s10052-014-3160-4",
    journal = "Eur. Phys. J. C",
    volume = "74",
    number = "11",
    pages = "3160",
    year = "2014"
}

@article{devuyst2024naturalintroductionfinetuning,
    author = "De Vuyst, Julian",
    title = "{A Natural Introduction to Fine-Tuning}",
    eprint = "2012.05617",
    doi = {10.48550/arXiv.2012.05617},
    archivePrefix = "arXiv",
    primaryClass = "physics.hist-ph",
    month = "12",
    year = "2020"
}

@article{Weinberg_2015,
    author = "Weinberg, David H. and Bullock, James S. and Governato, Fabio and Kuzio de Naray, Rachel and Peter, Annika H. G.",
    title = "{Cold dark matter: controversies on small scales}",
    eprint = "1306.0913",
    archivePrefix = "arXiv",
    primaryClass = "astro-ph.CO",
    doi = "10.1073/pnas.1308716112",
    journal = "Proc. Nat. Acad. Sci.",
    volume = "112",
    pages = "12249--12255",
    year = "2015"
}

@article{Riess_1998,
    author = "Riess, Adam G. and others",
    collaboration = "Supernova Search Team",
    title = "{Observational evidence from supernovae for an accelerating universe and a cosmological constant}",
    eprint = "astro-ph/9805201",
    archivePrefix = "arXiv",
    doi = "10.1086/300499",
    journal = "Astron. J.",
    volume = "116",
    pages = "1009--1038",
    year = "1998"
}

@article{sdsscollaboration2025nineteenthdatareleasesloan,
    author = "Pallathadka, Gautham Adamane and others",
    collaboration = "SDSS",
    title = "{The Nineteenth Data Release of the Sloan Digital Sky Survey}",
    eprint = "2507.07093",
    doi = {10.48550/arXiv.2507.07093},
    archivePrefix = "arXiv",
    primaryClass = "astro-ph.GA",
    month = "7",
    year = "2025"
}

@article{Plank_2018,
    author = "Aghanim, N. and others",
    collaboration = "Planck",
    title = "{Planck 2018 results. VI. Cosmological parameters}",
    eprint = "1807.06209",
    archivePrefix = "arXiv",
    primaryClass = "astro-ph.CO",
    doi = "10.1051/0004-6361/201833910",
    journal = "Astron. Astrophys.",
    volume = "641",
    pages = "A6",
    year = "2020"
}

@article{Zlatev:1998tr,
    author = "Zlatev, Ivaylo and Wang, Li-Min and Steinhardt, Paul J.",
    title = "{Quintessence, cosmic coincidence, and the cosmological constant}",
    eprint = "astro-ph/9807002",
    archivePrefix = "arXiv",
    doi = "10.1103/PhysRevLett.82.896",
    journal = "Phys. Rev. Lett.",
    volume = "82",
    pages = "896--899",
    year = "1999"
}

@article{Hogg:2020ktc,
    author = "Hogg, Natalie B. and Martinelli, Matteo and Nesseris, Savvas",
    title = "{Constraints on the distance duality relation with standard sirens}",
    eprint = "2007.14335",
    archivePrefix = "arXiv",
    primaryClass = "astro-ph.CO",
    reportNumber = "IFT-UAM/CSIC-20-114",
    doi = "10.1088/1475-7516/2020/12/019",
    journal = "JCAP",
    volume = "12",
    pages = "019",
    year = "2020"
}

@article{Mangano:2005cc,
    author = "Mangano, Gianpiero and Miele, Gennaro and Pastor, Sergio and Pinto, Teguayco and Pisanti, Ofelia and Serpico, Pasquale D.",
    title = "{Relic neutrino decoupling including flavor oscillations}",
    eprint = "hep-ph/0506164",
    archivePrefix = "arXiv",
    reportNumber = "DSF-16-2005, IFIC-05-17, MPP-2005-36",
    doi = "10.1016/j.nuclphysb.2005.09.041",
    journal = "Nucl. Phys. B",
    volume = "729",
    pages = "221--234",
    year = "2005"
}

@article{Fixsen:2009ug,
    author = "Fixsen, D. J.",
    title = "{The Temperature of the Cosmic Microwave Background}",
    eprint = "0911.1955",
    archivePrefix = "arXiv",
    primaryClass = "astro-ph.CO",
    doi = "10.1088/0004-637X/707/2/916",
    journal = "Astrophys. J.",
    volume = "707",
    pages = "916--920",
    year = "2009"
}

@article{Euclid:2020ojp,
    author = "Martinelli, M. and others",
    collaboration = "Euclid",
    title = "{Euclid: Forecast constraints on the cosmic distance duality relation with complementary external probes}",
    eprint = "2007.16153",
    archivePrefix = "arXiv",
    primaryClass = "astro-ph.CO",
    reportNumber = "IFT-UAM/CSIC-20-117",
    doi = "10.1051/0004-6361/202039078",
    journal = "Astron. Astrophys.",
    volume = "644",
    pages = "A80",
    year = "2020"
}

@article{Cutler:2009qv,
    author = "Cutler, Curt and Holz, Daniel E.",
    title = "{Ultra-high precision cosmology from gravitational waves}",
    eprint = "0906.3752",
    archivePrefix = "arXiv",
    primaryClass = "astro-ph.CO",
    doi = "10.1103/PhysRevD.80.104009",
    journal = "Phys. Rev. D",
    volume = "80",
    pages = "104009",
    year = "2009"
}

\newpage
\begin{appendix}
In this appendix we provide the numerical details for the current observational datsets employed to constrain the $\Lambda$CDM, GREA and modified GREA.
In particular, we report the cosmic chronometer data and the CMB compressed informations.
\section{Cosmic Chronometer data (CC)}
\label{ap:CC}
    The cosmic chronometers provide a direct, model-independent measurement of the Hubble paramater $H(z)$ based on the differential ages evolution of passively evolving galaxies.
    The data used in this work are taken from \citep{Arjona_2022} and are reported in \autoref{table:CCdata}, in units of $\text{km}$ $\text{s}^{-1}$ $\text{Mpc}^{-1}$.
    We assume that the measures of $H(z)$ are uncorrelated and we assume a diagonal covariance matrix.
    \begingroup 
    \setlength{\tabcolsep}{10pt}
    \renewcommand{\arraystretch}{1.5} 
    \begin{table}[H]
    \centering
    \begin{minipage}{0.45\textwidth}
    \centering
    \begin{tabular}{ c c c}  
    Redshift ($z$) & $H(z)$ & $\sigma_H$ \\
    \hline \hline
    0.07 & 69.0 & 19.6 \\
    0.09 & 69.0 & 12.0 \\
    0.12 & 68.6 & 26.2 \\
    0.17 & 83.0 & 8.0 \\
    0.179 & 75.0 & 4.0 \\
    0.199 & 75.0 & 5.0 \\
    0.2 & 72.9 & 29.6 \\
    0.27 & 77.0 & 14.0 \\
    0.28 & 88.8 & 36.6 \\
    0.35 & 82.7 & 8.4 \\
    0.352 & 83.0 & 14.0 \\
    0.3802 & 83.0 & 13.5 \\
    0.4 & 95.0 & 17.0 \\
    0.4004 & 77.0 & 10.2 \\
    0.4247 & 87.1 & 11.2 \\
    0.44 & 82.6 & 7.8 \\
    0.44497 & 92.8 & 12.9 \\
    0.4783 & 80.9 & 9.0 \\
    \hline \hline
    \end{tabular}
    \end{minipage}
    \begin{minipage}{0.45\textwidth}
    \centering
    \begin{tabular}{ c c c}  
    Redshift ($z$) & $H(z)$ & $\sigma_H$ \\
    \hline \hline
    0.48 & 97.0 & 62.0 \\
    0.57 & 96.8 & 3.4 \\
    0.593 & 104.0 & 13.0 \\
    0.60 & 87.9 & 6.1 \\
    0.68 & 92.0 & 8.0 \\
    0.73 & 97.3 & 7.0 \\
    0.781 & 105.0 & 12.0 \\
    0.875 & 125.0 & 17.0 \\
    0.88 & 90.0 & 40.0 \\
    0.9 & 117.0 & 23.0 \\
    1.037 & 154.0 & 20.0 \\
    1.3 & 168.0 & 17.0 \\
    1.363 & 160.0 & 33.6 \\
    1.43 & 177.0 & 18.0 \\
    1.53 & 140.0 & 14.0 \\
    1.75 & 202.0 & 40.0 \\
    1.965 & 186.5 & 50.4 \\
    2.34 & 222.0 & 7.0 \\
    \hline \hline
    \end{tabular}
    \end{minipage}\\
    \caption{Data of $H(z)$ in units of km/s/Mpc, with relative uncertainty \citep{Arjona_2022}.}
    \label{table:CCdata}
    \end{table}
    \endgroup

\section{Compressend Cosmic Microwave Background data (CMB)}
\label{ap:CMB}
    To incorporate early-Universe constraints and calibrate the sound horizon $r_d$, we use the compressed CMB parameters derived from the Plank and ACT (P-ACT) joit analysis presented in \citet{calderon2025constraininggreaalternativetheory}.
    The compressed data vector, $\mathbf{D}$:
    \begin{equation} 
    \label{eq:CMBdata}
    \mathbf{D} = \begin{pmatrix}
    100\,\theta^* \\
    \omega_b \\
    \omega_{bc}
    \end{pmatrix} = \begin{pmatrix}
    1.04165 \\
    0.0225046 \\
    0.141640
    \end{pmatrix},
    \end{equation}
where $\theta^*$ is the angular size of the size horizon at recombination, $\omega_b = \Omega_b h^2$ is the physical baryon density, and $\omega_{bc}=\Omega_{bc}h^2$ is the total physical matter density. 
The covariance matrix, $\mathbf{\Sigma}$, is:
    \begin{equation}
    \label{eq:CMBcovmat}
    \mathbf{\Sigma} = \begin{pmatrix}
    6.64717 \times 10^{-8} & 2.71701 \times 10^{-9} & -3.92400 \times 10^{-8} \\
    2.71701 \times 10^{-9} & 1.20267 \times 10^{-8} & -3.49601 \times 10^{-8} \\
    -3.92400 \times 10^{-8} & -3.49601 \times 10^{-8} & 1.11695 \times 10^{-6}
    \end{pmatrix}.
    \end{equation}

\end{appendix}

\end{document}